\documentclass[conference,compsoc]{IEEEtran}
\IEEEoverridecommandlockouts

\usepackage{graphics}
\usepackage[font=small]{subcaption}
\usepackage[font=small]{caption}
\usepackage{booktabs}
\usepackage{multirow}
\usepackage{amsthm}
\usepackage{amssymb}
\usepackage{amsmath}
\usepackage{dsfont}
\usepackage[ruled,vlined,linesnumbered]{algorithm2e}
\usepackage{mathtools}
\usepackage{xfrac}
\usepackage{float}
\usepackage{enumitem}
\usepackage{pgfplots}
\usepackage{tikz, tikzscale}
\pgfplotsset{compat=1.17}
\usepackage[most]{tcolorbox}
\usepackage{hyperref}
\usepackage{xurl}
\usepackage[nameinlink,capitalize,noabbrev]{cleveref}
\usepackage{titletoc}
\usepackage{cite}

\newtheorem{theorem}{Theorem}
\newtheorem{lemma}[theorem]{Lemma}
\newtheorem{corollary}[theorem]{Corollary}
\theoremstyle{definition}
\newtheorem{definition}[theorem]{Definition}

\newcommand{\boldparagraph}[1]{\noindent\textbf{#1}\mbox{}\quad\xspace}
\newcommand{\eps}{\varepsilon}

\makeatletter
\let\oldnl\nl%
\newcommand{\nonl}{\renewcommand{\nl}{\let\nl\oldnl}}%
\makeatother

\newcommand{\changed}[1]{{%
  \setlength{\fboxsep}{1pt}%
  \ifmmode
    \colorbox{yellow!30}{$#1$}%
  \else
    \colorbox{yellow!30}{#1}%
  \fi
}}
\def\HiLi{\leavevmode\rlap{\hbox to \hsize{\color{yellow!30}\leaders\hrule height .8\baselineskip depth .5ex\hfill}}}

\begin{document}

\title{
    MammothDP: Private Boosted Decision Trees Ready for Trusted Hardware and Leakage-free Hyperparameters
    \thanks{The research received funding from the BMFTR in the AnoMed project. Funded by the European Union – NextGenerationEU.}
}

 \author{
     \IEEEauthorblockN{Moritz Kirschte\IEEEauthorrefmark{1}, Thorsten Peinemann\IEEEauthorrefmark{1}, Kari Kostiainen\IEEEauthorrefmark{2}, \\ Jan Wichelmann\IEEEauthorrefmark{1}, Thomas Eisenbarth\IEEEauthorrefmark{1}, and Esfandiar Mohammadi\IEEEauthorrefmark{1}}
     \IEEEauthorblockA{\IEEEauthorrefmark{1}University of Luebeck\\
     Lübeck, Germany \\
     \{m.kirschte, t.peinemann, j.wichelmann, thomas.eisenbarth, esfandiar.mohammadi\}@uni-luebeck.de}
     \IEEEauthorblockA{\IEEEauthorrefmark{2}ETH Zurich\\
     Zurich, Switzerland \\
     kari.kostiainen@inf.ethz.ch}
 }

\maketitle
\thispagestyle{plain}
\pagestyle{plain}

\begin{abstract}
Data spaces like the European Health Data Space (EHDS) require the protection of sensitive data in studies during the data's processing and the study's release alike.
In this work, we investigate how to implement this mandate using established trusted execution environments (TEEs) and timing side-channel hardening to protect the processing even under a compromised server, and the state-of-the-art differential privacy (DP) notion to protect the release of studies on tabular data.
This combined setup could seem to complete the picture, but in fact raises unexpected security challenges that we mitigate by introducing MammothDP: a significant milestone towards a holistic machine learning system based on the recent gradient boosted decision tree learner S-BDT that is provably DP if used within a privacy filter, timing side-channel hardened, and deployed on a TEE. To achieve a strong privacy-utility tradeoff, we introduce two novel components: an early stopping criterion that adaptively selects the number of trees without any privacy overhead, and an improved rollback protection that is no longer vulnerable under timing side-channel attacks.
\end{abstract}

\section{Introduction}

Data spaces \cite{franklin2005databases} are starting to be adapted on a larger scale, e.g., at the EU level \cite{EUDataSpace} with the European Health Data Space (EHDS) \cite{EHDS} serving as a prominent example. Besides classical goals like high data utility or low computational overhead, security and privacy of data are stressed as an important data management challenge in data spaces~\cite{halevy2006principles}. This view is reflected in the EHDS regulation that mandates that studies on raw sensitive data should be processed inside a so-called \emph{secure processing environment} \cite[Art 73]{EHDS} and then released in an \emph{anonymized format} \cite[Art 66(2)]{EHDS}.

In this work, we investigate how to implement this mandate, i.e., how a data space can be designed that protects the data's \emph{processing} and study's \emph{release} alike. As a technical backbone, we use the established trusted execution environment (TEE) \cite{inteltdx,amdsev} and timing side-channel hardening \cite{ctimplementation1, ctimplementation2, xgboostontrustedhw, svmandmoreontrustedhw} to protect the \emph{processing} even under a compromised server, and the state-of-the-art differential privacy (DP) notion \cite{DBLP:conf/tcc/DworkMNS06} to protect the \emph{release} of studies on tabular data.\footnote{Other system setups are also plausible, like replacing TEEs with multi-party computation (MPC). Due to MPC's considerably high computational and communication overhead, we opt for a TEE-based DP setup as we find it more realistic and discuss an MPC-based approach in \Cref{sec:nontee_relatedwork}.} Based on the mandate and available technical solutions, we derive the following (strict) threat model:

\subsubsection*{Data Space Threat Model}
The attacker's goal is to infer which of two neighboring datasets $D, D'$ was used to release the study. To deploy a DP mechanism inside a TEE, we need to assume an attacker who (i) has the background knowledge to supply the release with $D,D'$ (\emph{typical DP assumption}), (ii) does not know what is processed inside the TEE but can rollback processes (\emph{typical TEE assumption}), and (iii) exploits data-dependent timing differences at the binary code level, including control flow and memory access (\emph{many typical timing side-channel attacks}).

\subsubsection*{Problem Statement \& Approach}
Deploying a system within this threat model that combines a \emph{processing} solution (TEE \& timing side-channel hardening) with a \emph{release} solution (DP) could seem to complete the picture, but in fact, raises unexpected security challenges that are pointed out in \Cref{sec:prob_stat} and go beyond the established constant-time programming and rollback protection~\cite{parno2011memoir,matetic2017rote,angel2023nimble,niu2022narrator,raoul2016adriadne,elephantdp}.
Deploying a system that mitigates these challenges can finally comply with our threat model but does not automatically imply a strong privacy-utility tradeoff. In fact, we point out in \Cref{sec:prob_stat} one major privacy-utility hurdle: common DP hyperparameter tuning is too privacy-costly.

We mitigate these security challenges and the privacy-utility hurdle with \emph{MammothDP}: a significant milestone towards a holistic machine learning system for tabular data that is provably DP if used within a privacy filter~\cite{rogers2016privacy,feldman2020individual,lecuyer2021practical}, timing side-channel hardened, deployed on a TEE, and tunes hyperparameters without excess privacy costs. 
To achieve a strong privacy-utility tradeoff, we use novel components highlighted in \Cref{sec:contrib} (``Our Contributions'').

\subsection{Detailed Problem Statement: Security Challenges of MammothDP}
\label{sec:prob_stat}
Releasing a study within a data space can come in many notions. \emph{MammothDP} focuses on releasing a gradient boosted decision tree (GBDT) ensemble, which fits for the following reasons: i) GBDTs are a common technique to learn features from data, ii) the recent DP variant of a GBDT learner \emph{S-BDT} \cite{sbdt} exhibits a strong utility-privacy tradeoff (incl. $\eps \le 0.1$) that fits our strict threat model, and iii) GBDTs offer two properties useful to address challenge P4: a strong theoretical basis and iterative training.

\subsubsection*{P1: DP Guarantees Deplete under Timing Side-channel Leakage}
\begin{figure}[!ht]
    \centering
    \includegraphics[width=\linewidth]{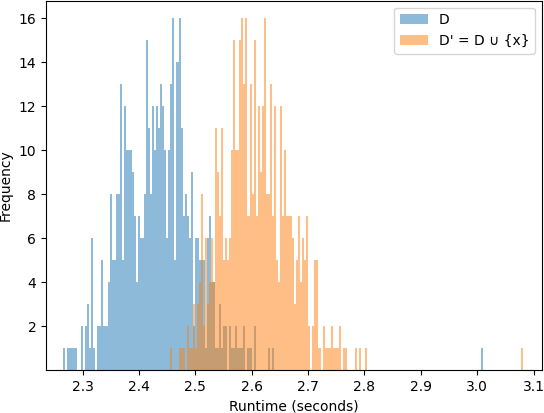}
    \caption{
    \textbf{Simplified timing attack on S-BDT: A single training point measurably impacts the runtime, although ($\varepsilon=0.1$)-DP holds}: the adversarial advantage is TPR=0.29@FPR=0.002 at threshold $\tau=2.64\,\mathrm{s}$.
    Setup: $500$ S-BDT models with $T=30{,}000$ trees of depth $d=8$ are each trained on a random subset $D$ of size 3 of Abalone \cite{misc_abalone_1} and similarly for $D'$.
    The metric TPR@FPR=0.002 quantifies the true positive rate (TPR) at a very low false positive rate (FPR) \cite{carlini2022membership}.
    }
    \label{fig:attack}
\end{figure}

A DP training algorithm \cite{DBLP:conf/tcc/DworkMNS06} ensures that an attacker can only deduce with a bounded probability from the output whether any particular data point was a member of the training. However, DP only bounds the privacy leakage of the output, not of the processing steps, such as timing leakage. As a consequence for MammothDP, although DP training algorithms like S-BDT \cite{sbdt} reach strong privacy guarantees (i.e., $\eps \le 0.1$), these guarantees become meaningless in a deployed system if it is easier for an attacker to deduce the membership of a data point not from the output of the model but by compromising the security guarantees that the TEE is supposed to provide. 
\Cref{fig:attack} illustrates a simple timing attack in which simply observing the runtime, which an adversary could measure in a TEE setup, allows the adversary to completely erode a strong $\eps=0.1$ DP guarantee.
More timing attacks are presented in the literature \cite{DBLP:conf/ccs/BulckPS18,DBLP:conf/ccs/WilkeS024} which also specifically target DP \cite{timing2025ratliff,jin2022we}.

\begin{tcolorbox}[colback=blue!5!white,colframe=blue!30!black,title=Challenge P1]
    Timing side-channels can break the DP guarantee.
\end{tcolorbox}
\begin{tcolorbox}[colback=orange!5!white,colframe=orange!70!black,title=Our Solution]
    For \emph{MammothDP}, we implemented a constant-time version of S-BDT and verified the robustness of the compiled binary against system-level attackers.
\end{tcolorbox}

\subsubsection*{P2: DP's Threat Model Incompatible With Timing Side-channel Hardening}
Classical timing side-channel hardening counters timing attacks by ensuring that control flow, memory access patterns, and instruction runtimes are the same for all input datasets. In a crypto context, FACCT \cite{facct} has shown that sampling noise, which is required for DP, though probabilistic, can also be constant-time. However, within the threat model of DP, timing side-channel hardening faces a new security challenge: \emph{differing dataset sizes}.

DP hides from an attacker whether an $n$-sized dataset $D$ or a $(n+1)$-sized dataset $D'$ was used for model training. Per definition of DP, the attacker can choose the neighboring datasets $D,D'$. Thus, a successful timing side-channel hardening has to ensure that the runtime of $D$ and $D'$ is always the same. 
Provably, such a side-channel hardening is impossible for many ML algorithms: training a model on $D'$ will in expectation always take longer than training on $D$ as the training algorithm has to loop over one additional data point.

A simple solution like padding the dataset to a fixed size introduces not only considerable overhead but still fails since an attacker can just choose a dataset with twice that size. Another solution is to just exchange the DP definition to bounded DP where neighboring datasets are of the same size. However, the leakage does not go away if we just exchange the privacy definition and remove the capability of the attacker to probe with differing dataset sizes.

\begin{tcolorbox}[colback=blue!5!white,colframe=blue!30!black,title=Challenge P2]
    No system can fully prevent timing side-channel leakage in the DP threat model.
\end{tcolorbox}
\begin{tcolorbox}[colback=orange!5!white,colframe=orange!70!black,title=Our Solution]
    For \emph{MammothDP}, we upper-bound the timing leakage that occurs if S-BDT runs on $D$ versus $D'$ and prove so-called timing privacy \cite{timingprivacy}.
\end{tcolorbox}

\subsubsection*{P3: Rollback Attacks on TEEs Erode DP Guarantees and the Previous Defense \emph{ElephantDP} Fails in our Threat Model, i.e. under Timing Side-channel Attacks}
TEEs guarantee data confidentiality and execution integrity. However, a TEE does not guarantee state continuity and is thus vulnerable to so-called rollback attacks \cite{parno2011memoir,matetic2017rote} that revert the TEE's persistent state to a previous version. In our threat model, these attacks seriously erode the DP guarantee \cite{elephantdp}: after reverting the TEE, the TEE reruns the same DP query but with freshly drawn noise, which allows an adversary to erode the DP guarantee by averaging out the noise.

The state-of-the-art \emph{ElephantDP} \cite{elephantdp} protects against rollback attacks, but only in the absence of timing side-channel attacks. In fact, they choose a design that fails under timing side-channel attacks, as they limit the release and not the computation of a query. Thus, an adversary can use a rollback attack to rerun the DP query multiple times and observe the runtime to accumulate timing leakage (cf. \Cref{fig:attack}), which erodes the DP guarantee (cf. P1 \& P2).
\begin{tcolorbox}[colback=blue!5!white,colframe=blue!30!black,title=Challenge P3]
  The state-of-the-art rollback protection \emph{ElephantDP} relies on a design that is vulnerable in our threat model, i.e., under timing side-channel attacks.
\end{tcolorbox}
\begin{tcolorbox}[colback=orange!5!white,colframe=orange!70!black,title=Our Solution]
    We propose a new design that mitigates such attacks by limiting the computation and not the release.
\end{tcolorbox}

\subsubsection*{P4: Common DP Hyperparameter-tuning is too costly in terms of $\eps$ for S-BDT's Strict Privacy Guarantees ($\eps \le 0.1$)}
For deployment, a challenge in our MammothDP system that even occurs without the TEE threat model is that outliers can noticeably influence the selection of hyperparameters \cite{papernot2022hyperparameter} and can thus leak privacy.
A common technique for privacy-preserving hyperparameter tuning is to try out numerous hyperparameters and select the best one, e.g., via differentially private voting. Such methods increase the leakage by a factor of 2 to 3 in terms of $\eps$ \cite{privateselection}. As we aim for strict privacy guarantees ($\eps \le 0.1$), these DP hyperparameter selection methods are prohibitively expensive. Other techniques solely rely on heuristics by choosing hyperparameters that turn out to work well but fail to adapt to the particular training task.
\begin{tcolorbox}[colback=blue!5!white,colframe=blue!30!black,title=Challenge P4]
  Prior DP hyperparameter tuning methods are too expensive for recent designs like S-BDT that target strict privacy guarantees ($\eps \le 0.1$).
\end{tcolorbox}
\begin{tcolorbox}[colback=orange!5!white,colframe=orange!70!black,title=Our Solution]
    We propose a new paradigm that combines two extremes: it not only costs no extra privacy but also remains adaptive to the learning task. For that, we only process intermediary-released noisy gradients (post-processing) to adaptively decide at each iteration whether to stop the training early.
\end{tcolorbox}

\subsection{Our Contributions}
\label{sec:contrib}

We introduce \emph{MammothDP}: a significant milestone towards a holistic TEE-based system of a timing side-channel hardened DP tree ensemble learner based on S-BDT with leakage-free adaptive hyperparameters and improved rollback protection.
\emph{MammothDP} addresses all problems (P1-P4) from the problem statement as follows:

\begin{enumerate}[align=left, leftmargin=*]
    \item[(P1 \& P2)] We provide a constant-time implementation of S-BDT\footnote{\url{https://github.com/kirschte/MammothDP/tree/main/tee-ready-image}} that is hardened against timing leakage due to
    \begin{itemize}[leftmargin=*]
        \item differing memory access patterns and control flow via adapting S-BDT to use bitmasking and to ensure that we always loop across all data points,
        \item differing runtime of instructions via tailoring the C++ library \emph{libfixedtimefixedpoint}\cite{libfixedtimefixedpoint} to S-BDT,
        \item differing noise sample runtimes via tailoring FACCT \cite{facct} to the Gaussian and Bernoulli noise in S-BDT, and
        \item differing data sizes within the threat model DP that lead to a differing number of loop iterations (and show so-called timing privacy \cite{timingprivacy}).
    \end{itemize}
    We validate our constant-time results using Microwalk~\cite{microwalk} and measure the computation overhead. 
    
    \item[(P3)] We identify that the state-of-the-art rollback protection ElephantDP \cite{elephantdp} is flawed in our threat model, i.e., under timing side-channel attacks. For MammothDP, we propose and evaluate a new design that mitigates such attacks by limiting the computation and not the release. Our new design is realized by adapting Memoir \cite{parno2011memoir} who use hash chains as a freshness tag.

    \item[(P4)] We introduce a novel early tree stopping technique for S-BDT that has leakage-free yet adaptively-selected hyperparameters at no extra privacy budget.
    For that, we only process intermediary-released noisy gradients (post-processing) to decide at each training iteration whether to stop the training early.
    Privacy-wise, our early stopping provably bounds the Rényi divergence and is, hence, compatible with provable DP bounds when used within a privacy filter.
    Utility-wise, we design our early stopping such that it triggers if we detect a possible point of bias-freeness. This insight is based on our theoretical foundation for regression based on the well-known division of the mean squared error (MSE) into bias and variance error, with the goal of minimizing the bias.
    We experimentally show that we still perform worse than S-BDT with privacy-leaky optimal hyperparameters but significantly outperform S-BDT with a statically chosen number of trees, especially for regression.
\end{enumerate}

\subsubsection*{Structure}
\Cref{sec:hardening,sec:rollback,,sec:earlytreestopping} cover our contributions addressing P1 \& P2, P3, and P4 respectively. Our evaluation is in \Cref{sec:evaluation}.

\section{Related Work}
\label{sec:related_work}

\subsubsection*{TEE Rollback Protection}
Many rollback protections exist \cite{parno2011memoir,matetic2017rote,angel2023nimble,niu2022narrator,raoul2016adriadne} with different state continuity modules (SCM), based on TPMs or distributed state-keeping systems. MammothDP and the recent ElephantDP \cite{elephantdp} abstract from a concrete SCM and focus on rollback protection within a DP attack model. However, ElephantDP ultimately relies on a design that is vulnerable in our threat model, i.e., under timing side-channel attacks. We discuss this limitation and how to build a better design in \Cref{sec:rollback}.

\subsubsection*{DP-ML With Leakage-Free Hyperparameters}
Current work mainly focuses on finding the best hyperparameter from a previously specified set of hyperparameter choices. In that, there are three approaches for leakage-free hyperparameters: via heuristics \cite{abadi2016deep,iyengar2019towards}, via separating the data into disjoint sets \cite{chaudhuri2011differentially,koskela2023practical}, or via spending extra privacy \cite{chaudhuri2013stability,abadi2016deep,privateselection,papernot2022hyperparameter}.
Heuristics are not adaptive to the training progress.
Spending extra privacy budget leads to a worse privacy-utility tradeoff. Dividing the dataset into disjoint sets implies using less data in training, which deteriorates the utility.
In this work, we chose a new paradigm that combines two extremes: it not only costs no extra privacy but goes one step beyond a heuristic by remaining adaptive to the learning task. For that, we reuse intermediary noisy gradients that the S-BDT training releases anyway (post-processing). Although it provably has a bounded privacy budget at each step, it adds additional complexity with the privacy filter to also reach provable DP bounds.

\subsubsection*{Hardening ML Algorithms for Trusted Hardware}
Prior works harden various machine learning algorithms against timing side-channels for trusted hardware. Law et al. \cite{xgboostontrustedhw} harden the XGBoost \cite{chen2016xgboost} algorithm against cache side-channels; Ohrimenko et al. \cite{svmandmoreontrustedhw} propose constant-time implementations for various ML algorithms, including decision trees and support vector machines. 
Sasy and Ohrimenko~\cite{sasy2019oblivious} propose data-oblivious subsampling that touches a subsample of size $B$ in $\Theta(B)$ rather than $\Theta(n)$ for a full scan. However, S-BDT still requires gradients and Hessians for all $n$ data points each round for individual privacy accounting. We therefore opt for easily verifiable linear scans (cf.\ \Cref{sec:data-dependent-memory-access}), especially as training a tree on $10^6$ points only takes $\approx 8$ seconds.

\section{Preliminaries}
\label{sec:prelims}

\subsection{Differential Privacy (DP)} 
DP \cite{DBLP:conf/tcc/DworkMNS06} is the de facto standard for provable privacy. Given a mechanism that has a data-dependent output, DP requires that the impact of a single data point on the output is limited and thus deniable. In line with S-BDT \cite{sbdt}, we consider the DP variant Rényi DP (RDP) in the unbounded definition, where the effect of adding or removing a single instance from a dataset on the output is analyzed.

\begin{definition}[Differential Privacy \cite{DBLP:conf/tcc/DworkMNS06}]\label{def:dp}
Let $2^{\mathcal{X}}$ be the set of all datasets over the universe of all rows $\mathcal{X}$, $\mathcal{R} \coloneqq \operatorname{RV}(X)$ be the set of all random variables over a given $X$, and $X \sim X'$ be two neighboring datasets where $X,X' \subseteq \mathcal{X}$ and $X' \coloneqq X \cup \{x\}$ for some $x \in \mathcal{X}$.\\
Given $\varepsilon \ge 0$ and $\delta \in [0,1]$, a randomized algorithm $\mathcal{M}\colon 2^{\mathcal{X}} \rightarrow \mathcal{R}$ satisfies $(\varepsilon, \delta)$-DP if, for any two neighboring datasets $X \sim X'$ and any $S \subseteq Y$:
$\Pr[M(X) \in S] \leq e^{\varepsilon} \cdot \Pr[M(X') \in S] + \delta$.
\end{definition}

\begin{definition}[Rényi Divergence \cite{renyi_divergence}]
    Given probability distributions $P,Q$ over $\mathcal{R}$ where $P(o)$ denotes the density of $P$ at $o$, the Rényi divergence of order $\alpha$ is
        $\textstyle \operatorname{D}_\alpha(P \parallel Q) \coloneqq \frac{1}{\alpha-1} \log \operatorname{E}_{o \sim Q} (\frac{P(o)}{Q(o)})^{\alpha}$.
\end{definition}

\begin{definition}[Rényi DP {\cite[Def. 4]{DBLP:journals/corr/Mironov17}}] \label{def:renyi_dp}
    A randomized mechanism $M\colon \mathcal{X} \mapsto \mathcal{R}$ satisfies $(\alpha, \rho(\alpha))$-RDP if for any neighboring datasets $X, X'$:
    $\operatorname{D}_\alpha(M(X) \parallel M(X')) \leq \rho(\alpha)$.
\end{definition}

With RDP, we can mathematically describe features like composition, filters \cite{feldman2020individual}, or subsampling \cite{pmlr-v97-zhu19c} and can convert it back to DP (cf. \Cref{sec:rdp_conversion}).
An important property of (R)DP is \emph{post-processing}: only working on the output of an (R)DP-mechanism does not increase the leakage.

\begin{lemma}[Postprocessing, {\cite[Sec. 3]{DBLP:journals/corr/Mironov17}}]
\label{lem:postprocessing}
Given a randomized mechanism $M$ and a randomized function $f\colon \mathcal{R} \rightarrow \mathcal{R}'$. If $M$ is $(\alpha,\rho(\alpha))$-RDP, then $f \circ M$ is also $(\alpha,\rho(\alpha))$-RDP.
\end{lemma}

\subsection{Gradient Boosted Decision Trees (GBDT)}
GBDTs \cite{friedman2001greedy} learn a sequence of decision trees $f \coloneqq (f^{(1)},\dots,f^{(T)})$ by iteratively correcting errors of prior trees. Let $X = \{(x_1, y_1), \ldots,\allowbreak (x_n, y_n)\} \subseteq \mathbb{R}^{m} \times \mathbb{R}$ denote a labeled dataset with $n$ data points and $m$ features. For simplicity, we denote $y$ as a label, although a regression target applies equally.
A tree ensemble $f$ minimizes
\begin{align}
    \textstyle \mathcal{L}^{(T)}(f, X) = \sum_{(x_i,y_i) \in X} l(\phi_f^{(T)}(x_i), y_i) + \sum_{t=1}^T \Omega(f^{(t)})
\end{align}
where $l$ is a twice-differentiable convex loss function, e.g., squared error or binary cross-entropy, and $\Omega(f^{(t)}) = \sfrac{1}{2}\lambda \lVert V^{(t)} \rVert^2$ is a regularization term on the vector of leaves $V^{(t)} = \operatorname{Leaves}(f^{(t)})$.

The ensemble prediction of a data point $x$ is
\begin{align}
    \label{eq:bdt_pred}\textstyle\tilde{y}^{(T)} \coloneqq \phi_f^{(T)}(x) = f^{(0)} + \eta \sum_{t=1}^T f^{(t)}(x)
\end{align}
which sums the prediction of each tree $f^{(t)}(x)$ scaled by learning rate $\eta$ and the init score $f^{(0)}$ which is constant for all $x$. Optionally, $f^{(t)}(x)$ is also clipped between $-\beta$ and $\beta$.

Each tree $f^{(t)}$ recursively predicts from a root node to the leaves: Each node has a split criterion that decides, based on the attributes of a data point $x$, whether the prediction of $x$ should continue with the left or right child. This process is repeated recursively until the maximal tree depth is reached. Then, the value of the leaf where $x$ landed, i.e., $f_j^{(t)}(x)$, is returned as the prediction. The ensemble prediction is then a learning-rate-weighted sum of tree predictions.

For training, we consider a variant of GBDTs that chooses their split criterion uniformly at random \cite{bojarski2014differentially} and only trains each leaf value.
For that, the learner XGBoost~\cite{chen2016xgboost} proposes Newton boosting, a second-order approximation of the loss function with gradient $g(\Tilde{y}_i,y_i) = \sfrac{\partial}{\partial \Tilde{y}_i} l(\Tilde{y}_i, y_i)$ and Hessian $h(\Tilde{y}_i,y_i) = \sfrac{\partial^2}{\partial \Tilde{y}_i^2} l(\Tilde{y}_i, y_i)$:
\begin{align}
    &\nonumber\textstyle\phantom{\approx~} l(\tilde{y}_i^{(t)}, y_i) = l(\tilde{y}_i^{(t-1)} + f^{(t)}(x_i), y_i) \approx \\
    &\textstyle l(\tilde{y}_i^{(t-1)}, y_i) + g(\tilde{y}_i^{(t)},y_i) f^{(t)}(x_i) + \frac12 h(\tilde{y}_i^{(t)},y_i) {f^{(t)}}^2(x_i).
\end{align}
The leave value $f_j^{(t)}$ is derived with the Newton method as:
\begin{align}\label{eq:leaf_value}
\textstyle f_j^{(t)}(x) = V_j^{(t)}(x) = -\frac{\sum_{(x_i,y_i) \in I^{(\text{train)}}_{\text{Leaf}_j}} g(\phi_f^{(t)}(x_i),y_i)}{\sum_{(x_i,y_i) \in I^{(\text{train)}}_{\text{Leaf}_j}} h(\phi_f^{(t)}(x_i),y_i) + \lambda}
\end{align}
for any train data point $x \in I^{(\text{train)}}_{\text{Leaf}_j}$ landing in the $j$-th leaf $V^{(t)}_j$.

\Crefname{algorithm}{Alg}{Algs}
\begin{algorithm}[!ht]
    \small
    \newcommand{\single}[1]{\texttt{TrainSingleTree}}
    \caption{Training skeleton of \textsc{S-BDT}.\label{alg:sbdt_skeleton}}
    \SetKwFunction{RandomBinaryTree}{RandomBinaryTree}
    \SetKwFunction{PrivacyFilter}{PrivacyFilter}
    \SetKwFunction{PoissonSubsample}{PoissonSubsample}
    \SetKwFunction{NoisyAverageLabel}{NoisyAverageLabel}
    \SetKwFunction{NoisyAverageGradient}{NoisyAverageGradient}
    \SetKwFunction{NoisyAverageHessian}{NoisyAverageHessian}
    \SetKwFunction{NoisyClipGradSum}{NoisyClipGradSum}
    \SetKwFunction{NoisyClipHessSum}{NoisyClipHessSum}
    \SetKwFunction{UpdateGradient}{UpdateGradient}
    \SetKwFunction{UpdateHessian}{UpdateHessian}
    \SetKwFunction{DPInitScore}{DPInitScore}
    \SetKwFunction{PredictEnsemble}{PredictEnsemble}
    \SetKwFunction{LeafIndexOf}{LeafIndexOf}
    \KwIn{$T$: number of rounds, $D$: training data}
    $f = (f^{(0)})$\tcp*{ensemble incl. init score}
    \For(\tcp*[f]{train tree $f^{(t)}$}){$t = 1$ \KwSty{to} $T$}{
      $f^{(t)} =$ \RandomBinaryTree{$\dots$}\;
      $D =$ \PrivacyFilter{$D, \dots$}\label{lin:privacy-filter}\; 
      $S =$ \PoissonSubsample{$D, \dots$}\;
      $\forall j\colon I^{(\text{train)}}_{\text{Leaf}_j} = [\,]$\tcp*{initialize leaf array}
      \For{\KwSty{each} sample $(x_i,y_i)$ \KwSty{in} $S$}{
        $\hat{y}_i = \text{\PredictEnsemble{$f,x_i$}}$\tcp*{\Cref{alg:sbdt_prediction_skeleton}}
        $g_i = g(\hat{y}_i, y_i)$\tcp*{gradient of loss $l$}
        $h_i = h(\hat{y}_i, y_i)$\tcp*{Hessian of loss $l$}
        $v_i = f^{(t)}.\text{root}$\;
        \While{$v_i$ \KwSty{is not} a leaf}{
            \tcp{find leaf of $x_i$}
            \lIf{$v_i.\text{split\_criterion}(x_i)$}{
                $v_i = v_i.\text{child}.\text{left}$
            } \lElse{
                $v_i = v_i.\text{child}.\text{right}$
            }
        }
        Append $x$ to $I^{(\text{train)}}_{\text{Leaf}_j}$ via $j=$\,\LeafIndexOf($v_i$)\;
      }
      \For{\KwSty{each} leaf $v_j$ \KwSty{in} $f^{(t)}$}{
        \tcp{cf. \Cref{eq:leaf_value} with DP}
        $f_j^{(t)}(\cdot) \coloneqq \frac{\tilde g^{(t)}_j}{\tilde h^{(t)}_j} = \frac{\text{\NoisyClipGradSum{$I^{(\text{train)}}_{\text{Leaf}_j}$}}}{\text{\NoisyClipHessSum{$I^{(\text{train)}}_{\text{Leaf}_j}$}}}$\label{lin:dp-leaf-sum}\;
      }
      Append tree $f^{(t)}$ to ensemble $f$\;
    }
    \Return $f$\;
\end{algorithm}
\Crefname{algorithm}{Algorithm}{Algorithms}

\begin{algorithm}[!ht]
    \small
    \caption{Prediction skeleton of \textsc{S-BDT}.\label{alg:sbdt_prediction_skeleton}}
    \SetKwFunction{Length}{Length}
    \SetKwFunction{PoissonSubsample}{PoissonSubsample}
    \SetKwFunction{NoisyAverageLabel}{NoisyAverageLabel}
    \SetKwFunction{NoisyAverageGradient}{NoisyAverageGradient}
    \SetKwFunction{NoisyAverageHessian}{NoisyAverageHessian}
    \SetKwFunction{UpdateGradient}{UpdateGradient}
    \SetKwFunction{UpdateHessian}{UpdateHessian}
    \SetKwFunction{DPInitScore}{DPInitScore}
    \SetKwFunction{LeafIndexOf}{LeafIndexOf}
    \KwIn{$x$: data point, $f$: S-BDT ensemble, $\eta$: learning rate}
        $\hat{y} = f^{(0)}$\;
        \For{$t = 1$ \KwSty{to} $\Length{f}-1$}{
            $v = f^{(t)}.\text{root}$\;
            \While(\tcp*[f]{find leaf of $x$}){$v$ \KwSty{is not} a leaf}{
                \lIf{$v.\text{split\_criterion}(x)$}{
                    $v = v.\text{child}.\text{left}$
                }\lElse{
                    $v = v.\text{child}.\text{right}$
                }
            }
            $\hat{y} = \hat{y} + \eta \cdot f_j^{(t)}(x)$ with $j=\text{\LeafIndexOf{$v$}}$\;
        }
    \Return $\hat{y}$\tcp*{cf. \Cref{eq:bdt_pred}}
\end{algorithm}

\subsubsection{S-BDT}
The state-of-the-art S-BDT \cite{sbdt} is a GBDT learner that is $(\alpha, \rho(\alpha))$-RDP by choosing the splits uniformly at random and adding Gaussian noise to the leaves, i.e., to a clipped Hessian and a clipped gradient sum.
\Cref{alg:sbdt_skeleton,alg:sbdt_prediction_skeleton} are skeletons of S-BDT training and prediction with the full pseudocode of our constant-time S-BDT in \Cref{sec:ctsbdtpseudocode}.

\subsection{Trusted Execution Environment (TEE)}
TEE architectures such as Intel SGX~\cite{costan16} and Arm TrustZone~\cite{pinto2019demystifying} offer so-called \emph{enclaves} that are designed to isolate applications from the rest of the system, including the untrusted OS. Intel TDX~\cite{tdx-demystified}, AMD SEV~\cite{amdsev}, and Arm CCA~\cite{arm-cca} isolate entire virtual machines (VMs) from any untrusted code running on the same physical computing platform. All such TEEs have the same high-level goal: to allow confidential workloads to be executed on a computing platform that is trusted as little as possible (e.g., the OS, hypervisor, or administrator can be malicious).

Most TEEs offer an isolated execution environment using memory encryption. In the case of Intel’s SGX, the processor keeps track of the currently executing code and allows confidential data to be accessed only when the correct enclave is running. TEEs enable the processing of sensitive data even if the adversary controls the rest of the computing platform (data confidentiality). TEEs also allow third parties, such as data owners, to verify that the expected code is running inside the TEE before provisioning sensitive data to it (execution integrity). These guarantees provide an attractive alternative to secure computation using cryptography, as TEEs can process confidential data efficiently in plaintext without the need for a trusted party~\cite{popa2024confidential}.

\section{Constant-time S-BDT (P1 \& P2)}
\label{sec:hardening}

Deploying S-BDT inside a TEE does not suffice to protect its processing: timing side-channel attacks can still leak information via data-dependent timing behavior. These attacks exploit shared hardware features like caches \cite{cacheattack1, cacheattack2, cacheattack3, cacheattack4, cacheattack5, cacheattack6} and branch prediction units \cite{bpuattack1, bpuattack2}, leak information to a malicious hypervisor using the page table \cite{pagetableattack1, pagetableattack2} or track a TEE's control flow using other microarchitectural resources via single-stepping \cite{DBLP:conf/ccs/BulckPS18,DBLP:conf/ccs/WilkeS024}.

The most widely used timing side-channel protection is software hardening via constant-time programming \cite{ctimplementation1, ctimplementation2, xgboostontrustedhw, svmandmoreontrustedhw} that ensures no data-dependent behavior on the architectural execution-trace layer, i.e., no data-dependent memory access or control flow, assuming the processor introduces no additional timing leakage for a small subset of operations. In this work, we additionally protect against differing program runtime by proving so-called Timing DP, and remove operand-dependent instruction runtime.
In this section, we briefly describe each timing side-channel and our mitigation.

\subsection{Data-dependent Memory Access}
\label{sec:data-dependent-memory-access}

\boldparagraph{Attack.} Among many attacks that target cache side-channels \cite{cacheattack2, cacheattack3, cacheattack4, cacheattack5, cacheattack6}, we highlight an example attack based on the Prime+Probe cache monitoring technique \cite{primeprobe} which was improved to allow server operators to attack TEEs \cite{cacheattack1}. 
First, the attacker \emph{primes} the cache, i.e., accesses memory such that the entire cache is filled with data of the attacker process. The victim then executes code with access to memory addresses depending on sensitive data. The attacker \emph{probes} which of their cache lines got evicted, telling them which of the memory addresses the victim accessed and, therefore, leaking information about sensitive data.

\boldparagraph{Leaky Components in S-BDT.} 
The vanilla implementation of S-BDT does not hide the patterns of which data points are accessed in memory. It leaks for each data point the leaf that the data point belongs to, which depends on the public split criteria of the tree and the data point's sensitive attributes. As a consequence, this reveals the data point's specific attributes, or at least an imprecise range for those attributes.
Memory access only to data points with leftover privacy budget reveals data points that still have per-data-point privacy budget, disallowed in individual privacy accounting \cite{feldman2020individual}. Accessing only a training round's subsample is disallowed in privacy amplification by subsampling \cite{pmlr-v97-zhu19c}, because that assumes that the subsample remains hidden from an attacker.

\boldparagraph{Our Mitigation.} Our constant-time implementation loops over all training data points in a fixed order, independent of whether they belong to a given leaf, independent of the remaining individual privacy budget, or subsample inclusion.
Thus, each data point is always loaded from memory for computations on sensitive data. We use bitmasks to mask out data points (cf. \Cref{alg:bitmaskloop}, changes from S-BDT highlighted) that should not contribute to the result of a computation because they are not in the current subsample, have no privacy budget, or do not belong to a given leaf.

\begin{algorithm}[t!]
    \small
    \caption{Bitmasked subsample sum. \label{alg:bitmaskloop}}
    \SetKwFunction{Bitmask}{Bitmask}
    \SetKwFunction{PoissonSubsample}{PoissonSubsample}
    \KwIn{$D$: dataset}
    \HiLi{$\Bitmask = \PoissonSubsample(D, \dots)$}\;
    \For{each sample $s$ in \changed{$D$}}{
        $\dots$\\
        Append $s.\text{gradient} \changed{\cdot \Bitmask{s}}$ to $v.\text{gradients}$\;
        Append $s.\text{hessian} \changed{\cdot \Bitmask{s}}$ to $v.\text{hessians}$\;
      }
\end{algorithm}

\subsubsection*{Efficiency}
Scanning all $n$ data points costs $\Theta(n)$ per tree instead of $\Theta(B)$ for a subsample of size $B \approx \gamma \cdot n$ ($\gamma \leq 0.5$). Data-oblivious subsampling~\cite{sasy2019oblivious} touches a subsample of size $B$ in $\Theta(B)$ rather than $\Theta(n)$ for a full scan. However, S-BDT still requires gradients and Hessians for all $n$ data points each round for individual privacy accounting. We therefore opt for easily verifiable linear scans (cf.\ \Cref{sec:data-dependent-memory-access}), especially as training a tree on $10^6$ points only takes $\approx 8$ seconds.

\subsection{Data-dependent Control Flow} 
\label{sec:data-dependent-control-flow}
\boldparagraph{Attack.}
Single-stepping \cite{DBLP:conf/ccs/BulckPS18,DBLP:conf/ccs/WilkeS024} allows a server operator to precisely track the control flow of a TEE by sending hardware interrupts. These interrupts are served \emph{after} the currently executing instruction has completed, and take a variable amount of time depending on the instruction type and the microarchitectural state of the processor. This allows differentiating between program branches.

\boldparagraph{Leaky Components in S-BDT.}
Vanilla S-BDT's tree prediction evaluates the path from the tree root to the leaf reached by a data point, and an attacker can track this path and leaf by single-stepping. The particular leaf of a data point can leak sensitive information (cf. \Cref{sec:data-dependent-memory-access}).

\boldparagraph{Our Mitigation.}
Our constant-time version traverses the entire tree in a fixed order, independent of the data point. Instead of branching in traversal, we compute every split criterion in the entire tree and use a constant-time selection procedure (cf. \Cref{alg:constanttimeselection}) to select the correct leaf value.

\begin{algorithm}
    \small
    \caption{Constant-time prediction (call with root node). \label{alg:constanttimeselection}}
    \SetKwFunction{Gauss}{Gauss}
    \SetKwFunction{Clamp}{Clamp}
    \SetKwFunction{PredictTree}{PredictTree}
    \SetKwFunction{max}{max}
    \SetKwInOut{KwIn}{\color{white}\phantom{Input}} 
    \SetKwBlock{Server}{PredictTree(node $v$, data point $x$):}{}
    \nonl\Server{
    \If{$v$ is leaf}{
        \Return prediction of $v$\;
    }
    $\text{l} = \changed{\text{\PredictTree}(v.\text{child}.\text{left}, x)}$ \;
    $\text{r} = \changed{\text{\PredictTree}(v.\text{child}.\text{right}, x)}$ \;

    \Return \changed{$v.\text{split\_criterion}(x) \cdot l +  (1-v.\text{split\_criterion}(x)) \cdot r$}}
\end{algorithm}

\subsection{Data-dependent Runtime of Instructions}

\boldparagraph{Attack.}
Due to the complexity of floating-point numbers, modern processors can struggle to handle certain inputs efficiently and runtime measurements become a side-channel. For example, for subnormal operands, i.e., ones close to zero, multiplication can exceed 200 clock cycles on an Intel Core i7 processor, but takes 4 cycles for normal operands \cite{libfixedtimefixedpoint}.

\boldparagraph{Leaky Components in S-BDT.}
S-BDT repeatedly performs operations on sensitive gradients and Hessians or bitmasks used for masking out data not in a subsample.

\boldparagraph{Our Mitigation.}
We avoid ordinary instructions whenever an operand is sensitive. Instead, we use constant-time equivalents from the libftfp \cite{libfixedtimefixedpoint} math library for the affected instructions. 
libftfp ensures to execute every code path, produce every intermediary result instead of returning results early, use no data-dependent branching, loop with a fixed number of iterations, and use only basic integer operations that are expected to be constant-time.

\subsection{Leaking Noise Samples}

\boldparagraph{Attack.} 
Standard noise samplers often have timing side-channels. Samplers based on randomized trees, branch over the tree to then select a random sample from a leaf, thus leaking the branch decisions (e.g. by single-stepping, cf. \Cref{sec:data-dependent-control-flow}) leaks the random sample \cite{ConstantTimeDiscreteGaussianSampling}.  Samplers based on lookup-tables access memory of lookup-table entries that are used to generate the random sample, therefore leaking the memory access patterns (e.g. via cache attacks, cf. \Cref{sec:data-dependent-memory-access}) can leak information about the random sample \cite{groot2016flush}. 

\boldparagraph{Leaky Components in S-BDT.}
S-BDT uses randomness at three stages: two times Laplace noise to noise its DP initial score, multiple times Gaussian noise to noise the leaf values, and privacy amplification by Poisson subsampling uses Bernoulli noise to decide the inclusion of a data point in a training round. 

An attacker who is able to leak the Gaussian or Laplace noise samples can denoise a DP release. Privacy amplification by Poisson subsampling assumes the subsample remains hidden \cite{pmlr-v97-zhu19c}; thus, leaking the randomness that determined the subsample negates the privacy amplification.

\boldparagraph{Our Mitigation.} 
We replace S-BDT's Gaussian and Laplace samplers with FACCT~\cite{facct} (detailed in \Cref{sec:additional-preliminaries-facct}), a constant-time sampler for the discrete Gaussian $D_\sigma$ over the integers, and use $X/\sigma$ with $X \sim D_{\sigma}$, $\sigma = 2048$, in place of a standard normal sample.
On a target machine with finite precision, every implementable Gaussian sampler outputs values on a grid, and the best achievable distribution is $\mathcal{N}(0,1)$ rounded to that grid. As $\sigma$ grows, this rounded continuous Gaussian and our rescaled discrete Gaussian become statistically closer. In other words, our sampler approaches the limit of ``continuity'' that the finite precision allows, as does any real-world implementation of a continuous sampler.

We replace the Bernoulli sampler for Poisson subsampling with the constant-time implementation of FACCT \cite{facct}.

\subsubsection*{Implications}
Replacing Laplace noise with Gaussian noise in S-BDT's DP initial score yields the Gaussian mechanism instead of the Laplace mechanism; consequently, those releases satisfy approximate DP rather than pure DP, and we account for this $\delta$ accordingly. 
On the regression dataset \emph{Abalone} with small $\eps$, constant-time S-BDT shows a slight RMSE decrease over the vanilla version, due to the replacement. Since the initial score is only activated for regression, this impact is limited to the regression dataset.

For DP algorithms that use continuous noise distributions, the implication of running a DP algorithm on a real system with fixed precision is often abstracted away in the literature, and we are aware of the weaknesses.

\subsection{Leaking Input Length}

\boldparagraph{Attack.} In unbounded DP, neighboring datasets differ by exactly one data point. If the training runtime depends on the number of data points, an untrusted server operator observing the training runtime can distinguish the input lengths, revealing which dataset was processed and eroding the DP guarantee. Simple input padding to a fixed size does not suffice, since the attacker can choose a dataset that is much larger than that size.

\boldparagraph{Leaky Components in S-BDT.} Since vanilla S-BDT loops over the data points, its runtime can reveal the number of data points.

\boldparagraph{Our Mitigation.}
We make the observable runtime depend only on a DP release of the input length. For input length $l$ we compute a noisy input length $\hat{l} = l + Z + 2 \cdot \sigma_\text{input},  \qquad Z \sim \mathcal{N}(0,\sigma_\text{input}^2)$, and then 
pad data with random placeholder data, or truncate if necessary (cf. \Cref{alg:mammothdptimingprivate}), i.e. remove data points. We add a slack of $2 \cdot \sigma_\text{input}$ to likely avoid truncation. Afterwards, all S-BDT loops use the noisy length $l'$, so runtime observations reveal at most $l'$, rather than the exact input length $l$. Data owners perform this padding or truncation before the server executes S-BDT, and the server masks out placeholder data using bitmasks (cf. \Cref{sec:data-dependent-memory-access}).

For ease of analysis, we excluded an extra $(\varepsilon, \delta)$ from our experiments because the noisy input length costs computation but no utility, as an inaccurate input length only means looping over extra placeholder data, and the privacy cost stays tiny. For $(\varepsilon, \delta)=(0{.}001,10^{-11})$, a Gaussian with $\sigma_\text{input}=5013$ suffices and $\pm5{,}000$ iterations is negligible for datasets with $120{,}000$ points.

\begin{algorithm}
\small
\SetKwBlock{Client}{Each data owners:}{}
\SetKwBlock{Server}{Server:}{}
\caption{Noisy input length constant-time S-BDT \label{alg:mammothdptimingprivate}.}
\LinesNumbered
\SetAlgoNoLine
\SetAlgoNoEnd
\SetKwFunction{Bitmask}{Bitmask}
\nonl\Client{
    \KwIn{$D$: dataset}
    \SetKwInOut{KwIn}{\color{white}\phantom{Input}}
  \HiLi{$\tilde{l} = |D| + \mathcal{N}(0, \sigma_\text{input}^2) + 2 \cdot \sigma_\text{input}$}\;
  \HiLi{Add $\tilde{l} - |D|$ placeholder data or remove data from $D$}\;
  \HiLi{Generate \Bitmask of placeholder data}\;
}
\nonl\Server{
  Collect data \changed{and \Bitmask} from data owners\;
  Run \changed{\Bitmask masked} constant-time S-BDT\;
}
\end{algorithm}

\subsubsection{Timing Privacy}
We prove so-called Timing Privacy \cite{timingprivacy} in an idealized RAM model to validate our construction of \Cref{alg:mammothdptimingprivate}. Intuitively, Timing Privacy demands the randomized running time of a mechanism not to leak much more than what is already revealed by the output distribution.
The formal definition of Timing Privacy, the formal statement, and our proof are given in \Cref{sec:timing_privacy}.

\section{Improved TEE Rollback Protection (P3)}
\label{sec:rollback}

Rollback attacks present a severe and practical threat to TEE-based DP systems that run on untrusted computing platforms, such as cloud environments. Such systems only guarantee DP properties if \emph{(a)} any output released by the TEE is the result of a differentially private computation and \emph{(b)} the TEE has correctly enforced that there was sufficient \emph{privacy budget} available for the requested computation. In a simple rollback attack, the adversary requests the TEE to perform a DP computation. The TEE implementation checks that it has sufficient privacy budget, performs the computation, updates the remaining budget, seals its sensitive data and updated budget, and releases the computation result. In sealing, the TEE architecture derives an encryption key that is specific to each TEE application, which can then encrypt, decrypt, and authenticate data that it needs to store externally. After that, the adversary controlling the untrusted server requests the TEE to perform another DP computation, but instead of providing the latest seal to the TEE, it provides a previous version, which allows the adversary to reuse the privacy budget and thus violate DP guarantees.

\subsection{Rollback Protection Requirements}

We argue that a \emph{secure} and \emph{practical} rollback protection solution should meet the following four requirements. 

\boldparagraph{R1: Budget Integrity.} The main requirement is privacy budget integrity. The TEE implementation, realizing the DP algorithm, must always perform the mandatory privacy budget check using the correct (latest) budget value. This requires that the TEE implementation must only accept the latest sealed data when restoring its persistent state at the beginning of a new TEE execution. 

\boldparagraph{R2: Crash Tolerance.} Another common requirement for rollback protection is crash tolerance. After an unexpected crash, the TEE implementation should be restartable, and it can continue answering DP computation requests. In particular, the way the freshness-tag check is implemented should not introduce deadlocks. We note that more general availability guarantees are typically not possible for TEE systems where the adversary, who controls the untrusted computing platform, can start and stop TEE executions freely, and thus trivially prevent any TEE execution.

\boldparagraph{R3: Timing Side-channel Minimization.} Execution in modern TEE architectures is known to produce \emph{timing side-channel traces} that the adversary can use to infer secrets, such as sensitive training data. The TEE implementation should ensure that the adversary \emph{cannot} collect timing side-channel traces beyond the legitimate algorithm executions permitted by the privacy budget. We consider such side-channel minimization a complementary protection to TEE implementation hardening described in Section~\ref{sec:hardening}.

\boldparagraph{R4: Compatibility.} Finally, we argue that the rollback protection solution should be compatible with various state continuity modules. In particular, the rollback protection solution should work with both standard security chips, such as TPMs, and custom distributed state-keeping systems (e.g., research proposals like~\cite{angel2023nimble, niu2022narrator}).

\subsection{Limitations of Prior Work}\label{sec:priorwork_rollback}

ElephantDP~\cite{elephantdp} is a state-of-the-art rollback protection system for DP algorithms. Its design works as follows: when the TEE is started, it loads its persistent state from disk and compares a counter value used as the freshness tag to the SCM content. To evaluate a requested DP computation, the TEE implementation first checks that there is sufficient privacy budget available, then executes the computation, and updates the remaining privacy budget. After that, the TEE implementation writes its updated persistent state to disk. The state contains the latest computation result, the updated privacy budget, the computation request, and an incremented execution counter. Finally, the TEE writes this state to SCM. The design assumes that the SCM implementation is able to enforce the following rule: The written counter value must be exactly one higher than the previous value stored in the SCM. Otherwise, it reports failure.

However, this approach does not minimize timing side-channels (\textbf{P3}): the adversary can request a DP computation, collect timing side-channel traces from its execution, kill the TEE before anything is saved on disk, and repeat this process. In effect, the design creates a \emph{timing side-channel oracle} that the adversary can query with freely-chosen computation requests for an unlimited number of times. The root cause behind this problem is that, by updating the SCM after the DP computation, ElephantDP focuses on controlling the \emph{release} of DP computation results from the TEE, but it does not limit the actual execution of DP computations.

ElephantDP's design has compatibility limitations (\textbf{P4}): their rollback protection solution requires an SCM that compares the written value to the currently stored value. While such checks can be implemented in custom distributed systems like~\cite{angel2023nimble}, they are not available in all standard security chips like TPMs that support only a simple read and write interface to non-volatile memory (NVRAM).

\subsection{Overview of Our Solution}

\begin{figure}[t]
    \centering
    \includegraphics[width=\linewidth]{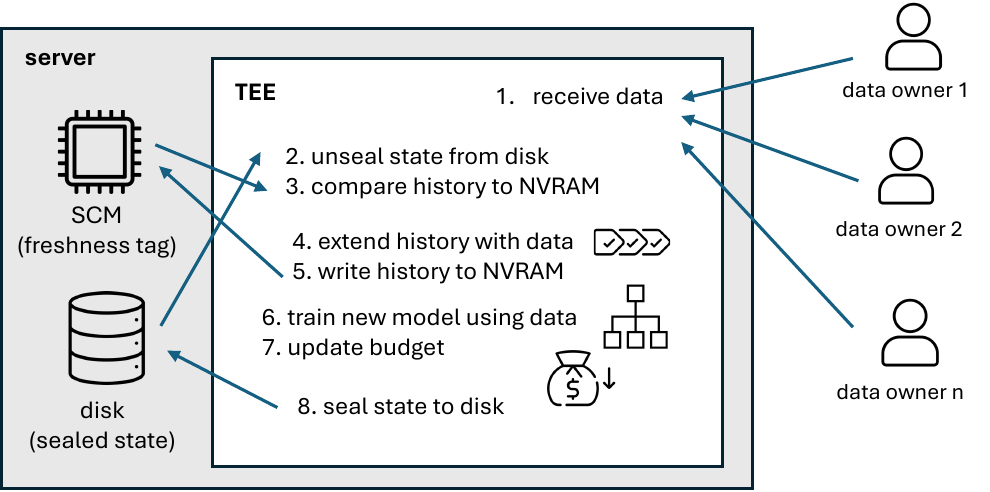}
    \caption{Overview of our rollback protection solution.}
    \label{fig:rollback-overview}
\end{figure}

We observe that the rollback protection requirements can be met by adapting a technique previously introduced by Parno et al.~in a paper called Memoir~\cite{parno2011memoir}. Parno's main idea is to use a \emph{hash chain} of all execution inputs as the freshness tag. This technique is compatible with various SCMs (\textbf{R4}) as it only requires a simple read and write interface. It also provides the basis for both budget integrity enforcement (\textbf{R1}) and crash tolerance (\textbf{R2}). To ensure timing side-channel minimization (\textbf{R3}), in the context of TEE-based DP implementations, new technical challenges must be solved.

An overview of our solution is shown in Figure~\ref{fig:rollback-overview}. Here, we illustrate the training phase of an ML model, but note that the model query phase would work similarly (data owners would be replaced by data analysts, training data would be replaced by queries, and model training would be replaced by model inference).

We initialize the system by collecting randomness from each data owner. During the first TEE execution, the TEE receives randomness shares from all the data owners and derives a pseudo-random seed based on the received shares. The TEE implementation then seals to disk its initial state, which contains the derived seed, an empty ML model, a full privacy budget value, and the value zero to represent an empty hash chain. The TEE also writes the zero value as a freshness tag to the SCM to complete the initialization.

Training iterations proceed as illustrated in Figure~\ref{fig:rollback-overview}. (1) Data owners send new training data to the TEE that (2) unseals its persistent state from disk and (3) checks that a freshness tag (initially zero) included in the seal matches the value stored in the SCM. If these checks pass, (4) the TEE then extends the (initially empty) hash chain with the input (training) data and (5) writes the extended chain value to the SCM. Note that the freshness tag is advanced to the next value before the actual DP computation takes place. This is essential to prevent timing side-channel oracles. (6) The model training algorithm can now be executed using the input (training) data, and (7) the TEE will update the privacy budget accordingly. Finally, (8) the TEE seals its updated persistent state (including the trained model, updated budget, and extended hash chain as the freshness tag) to disk. At this point in the process, the persistent state also matches the previously advanced freshness tag. The TEE is now ready to be closed down and wait for its next execution.

Additionally, the solution includes a recovery routine that re-executes the previous input (and only that input) after a crash during DP algorithm execution. During recovery, the TEE reads the previous state seal from the disk, extends the hash chain with the same (latest) input data, and only proceeds with the DP computation if the extended hash chain value matches the SCM content.

\subsubsection{Security Guarantees} Our approach preserves budget integrity (\textbf{R1}), because the TEE will only accept the latest seal or the previous seal with the same input. In both cases, the privacy budget is enforced correctly for the sensitive training data. This approach is also crash resilient (\textbf{R2}). If the DP algorithm execution crashes, we have a situation where the updated state seal is not created, and the previous state seal, still available on disk, no longer matches the extended hash chain (freshness tag) stored in the SCM. However, since the freshness tag is a deterministic hash chain, our recovery routine can safely allow the re-execution of the latest input and only that input. 

This also minimizes timing side-channels (\textbf{R3}). As the freshness tag is updated and written to the SCM \emph{before} the DP computation, the TEE implementation cannot serve as a timing side-channel oracle. The key difference to ElephantDP is that, in our solution, we focus on controlling DP computation inside the TEE, instead of controlling the release of DP computation results. To ensure that the re-executions of the latest input during the recovery routine do not benefit the adversary, we must carefully ensure that all the pseudorandom values used in the computation remain the same across executions. We achieve this by fixing the randomness seed during the first TEE execution. Furthermore, through code hardening (see Section~\ref{sec:hardening}), we ensure that there is no leakage from arithmetic operations and similar instructions known to exhibit microarchitectural leakage. Such protections are new to our rollback solution.

Finally, the approach is compliant (\textbf{R4}) with SCMs like TPMs that only support a simple read and write interface. We defer the details of our rollback protection design, its security analysis, and implementation notes to \Cref{sec:rollback-appendix}.

\section{Leakage-free Yet Adaptive Hyperparameters: Early Tree Stopping (P4)}
\label{sec:earlytreestopping}
A challenge in our MammothDP system that even occurs without a TEE is that outliers can noticeably influence the selection of hyperparameters \cite{papernot2022hyperparameter} and can thus leak privacy.
To solve this challenge, we propose a new paradigm that combines two extremes: it not only costs no extra privacy but also remains adaptive to the learning task.
For that, we fix all hyperparameters except the number of trees $t$. During each training iteration, we now only process intermediary-released noisy gradients (post-processing) to adaptively decide whether to stop the training early.

We thus assume that the number of trees is the most important hyperparameter which can cover up other suboptimally chosen hyperparameters and illustrate that on two examples: i) Decreasing parameters like the subsampling rate $\gamma$ or maximal tree depth $d$ results in a weaker tree. If each tree gets weaker, we can cover that up by using more trees. ii) If we have a larger dataset, we can increase the number of trees to reach a similar signal-to-noise ratio.

\subsubsection*{Structure} In \Cref{sec:earlystop_theory}, we show that the leaf values of a regression tree can approximate the direction of the bias error as part of the MSE metric. In \Cref{sec:earlystop_alg}, we show \textsc{EarlyStoppedSBDT} which stops the training early based on this insight that (noisy) leaf values can approximate the MSE error, and discuss its privacy implications in \Cref{sec:privacy_earlystopping}. 

\subsection{Theoretical Foundation for Regression}
\label{sec:earlystop_theory}

The mean squared error (MSE) is a common metric for regression predictors like GBDTs. For true label $y$ and predicted label $\tilde{y}$, it is: $\operatorname{MSE}(y,\tilde{y}) = \frac{1}{n}\sum_{i=1}^n (y_i - \tilde{y}_i)^2$. If each error $e_i \coloneqq y_i - \tilde{y}_i$ is an iid element of the error distribution $\mathcal{E}$, we can also write: $\operatorname{MSE}(\mathcal{E}) = \operatorname{E}[\mathcal{E}]^2$.
The MSE consists of a bias part $\operatorname{E}[\mathcal{E}]$ and a variance part $\operatorname{Var}(\mathcal{E})$ due to the relation $\operatorname{E}[\mathcal{E}]^2 = \operatorname{Var}(\mathcal{E}) + (\operatorname{E}[\mathcal{E}])^2$ for a random variable $\mathcal{E}$. The bias error (underfitting) is mostly caused by a systematic error in the learning, whereas the variance error (overfitting) is mostly caused by non-deterministic behavior of data which is not perfectly predictable. Data frequently consists of unlearnable white noise and a learnable correlated part, and as the white noise can not be fitted, variance error will remain. In contrast, the bias error can vanish if the correlated part is predicted well. Thus, a predictor that minimizes the bias is often also a good predictor with a small MSE.

As a proxy for stopping the tree training as soon as the bias is small, we determine the bias error difference quotient $\nabla B$: a discretized gradient that indicates how the bias error changes with each new tree. Formally, we have $\nabla B = \sum_{i=1}^n e_i^{(t)} - \sum_{i=1}^n e_i^{(t-1)}$ with $e_i^{(t)} \coloneqq y_i - \tilde{y}^{(t)}_i$ as the error term of the $t$-th tree on (test) input $(x_i,y_i)$. If $\nabla B = 0$, we have an extremum (i.e., maximum, minimum, or saddle point) for the bias error. A saddle point is not stable and will likely vanish if we train until the $(t+1)$-th tree. A maximum or minimum indicates that the bias that is unsigned has grown/fallen during training and now reaches a saturation point. If we assume that we have a useful configuration and thus make progress during training, reaching the first extremum is a good heuristic that the bias might have reached a small value.

We simplify the bias error difference quotient $\nabla B$:
\begin{align}
    &\nonumber\textstyle\nabla B = \sum_{i=1}^n e_i^{(t)} - \sum_{i=1}^n e_i^{(t-1)} \\
    &\nonumber\textstyle\phantom{\nabla B}= \sum_{i=1}^n y_i - \tilde{y}_i^{(t)} - \big(\sum_{i=1}^n y_i - \tilde{y}_i^{(t-1)}\big) \\
    &\nonumber\textstyle\phantom{\nabla B}= \sum_{i=1}^n \tilde{y}_i^{(t)} - \tilde{y}_i^{(t-1)}
    \intertext{by definition: $\tilde{y}_i^{(t)} = f^{(0)} + \eta \sum_{t'=1}^t f^{(t')}(x_i)$ (cf. \Cref{eq:bdt_pred}) with $f^{(t')}(x_i)$ as the $t'$-th tree prediction on (test) input $x_i$}
    &\nonumber\textstyle\phantom{\nabla B}= \eta\sum_{i=1}^n \big(\sum_{t'=1}^t f^{(t')}(x_i) - \sum_{t'=1}^{t-1} f^{(t')}(x_i)\big) \\
    &\textstyle\phantom{\nabla B}= \eta\sum_{i=1}^n f^{(t)}(x_i).
\end{align}

By definition of a leaf value $V_j^{(t)}(x)$ in \Cref{eq:leaf_value}, we have for any data point $x \in I_{\text{Leaf}_j}$ that ended up in the $j$-th leaf $V_j^{(t)}$:
\begin{align}
    \textstyle f_j^{(t)}(x) = V_j^{(t)}(x) = -\frac{\sum_{(x_i,y_i) \in I^{(\text{train)}}_{\text{Leaf}_j}} g(\phi_f^{(t)}(x_i),y_i)}{\sum_{(x_i,y_i) \in I^{(\text{train)}}_{\text{Leaf}_j}} h(\phi_f^{(t)}(x_i),y_i) + \lambda}.
\end{align}
As an approximation, we assume that the number of train and test data points that land in a leaf are roughly equal: $\lvert I^{(\text{train)}}_{\text{Leaf}_j} \vert \approx \lvert I_{\text{Leaf}_j} \rvert$.
Since for all $x \in I_{\text{Leaf}}$, $V_j^{(t)}(x)$ is the same, we simplify $\nabla B$ further:
\begin{align}
    &\nonumber\textstyle\eta\sum_{i=1}^n f^{(t)}(x_i)
    \textstyle= \eta\sum_{j=1}^{\lvert\operatorname{Leaves}(f^{(t)})\vert}\sum_{x \in I_{\text{Leaf}_j}} V_j^{(t)}(x)\\
    \intertext{if $\lambda=0$ and since for regression the Hessian $h^{(t)}(f^{(t)}(x_i),y_i) = 1$}
    &\nonumber\textstyle= -\eta\sum_{j=1}^{\lvert\operatorname{Leaves}(f^{(t)})\vert} \frac{\lvert I_{\text{Leaf}_j} \rvert}{\lvert I^{(\text{train)}}_{\text{Leaf}_j} \rvert} \sum\limits_{(x_i,y_i) \in I^{(\text{train)}}_{\text{Leaf}}} g(\phi^{(t)}(x_i),y_i)\\
    &\textstyle\approx -\eta\sum_{i=1}^n g(\phi^{(t)}(x_i),y_i).
\end{align}
Thus, we can conclude as a stopping criterion
\begin{align}
    \textstyle\label{eq:stop_crit}\nabla B \approx -\eta\sum_{i=1}^n g(\phi^{(t)}(x_i),y_i) \approx 0,
\end{align}
i.e., that if we stop the training when the sum of the training gradients is zero, the bias error difference quotient $\nabla B$ is approximately zero; thus, we probably reach a saturation point in the bias which may keep the learnable correlated component of the MSE small. Naturally, if the systematic ``bias'' error vanishes, only the variance error is left, which marks a good heuristic to stop early.

\subsection{Algorithm Overview}
\label{sec:earlystop_alg}

\Crefname{algorithm}{Alg}{Algs}
\begin{algorithm}[!ht]
    \small
    \newcommand{\single}[1]{\texttt{TrainSingleTree}}
    \caption{Training of \textsc{EarlyStoppedSBDT}.\label{alg:earlystop}}
    \SetKwFunction{RDPAccountant}{RDPAccountant}
    \SetKwFunction{TrainSingleTree}{TrainSingleTree}
    \SetKwFunction{DPInitScore}{DPInitScore}
    \KwIn{$T$: \changed{maximal} number of rounds, $D$: training data, $(r_1, r_2)$: noise weights for leaf value, $\sigma^2_{\text{leaf}}$: leaf noise scale, $d$: depths of trees, $g^*$: gradient clipping bound, \changed{$\sigma_{\text{CI}}$: cutoff multiplier}}
    $f = (f^{(0)})$)\label{alg:earlystop_l1}\tcp*{ensemble incl. init score}
    \HiLi{$\tau = \sigma_{\text{leaf}} \cdot g^* \cdot \sqrt{\sfrac{1}{2 r_1}} \cdot\sqrt{2^d}$}\tcp*{cutoff threshold}\label{alg:earlystop_l2}
    \HiLi{$(\Sigma g, \mathrm{dir}) = (0,\texttt{UNDEF})$}\label{alg:earlystop_l3}\;
    \For(\tcp*[f]{train tree $f^{(t)}$}){$t = 1$ \KwSty{to} $T$\label{alg:earlystop_l4}}{
      \tcp{\Cref{alg:sbdt_skeleton}: train $f^{(t)}$ and append to $f$}\label{alg:earlystop_l5}
      \HiLi{\tcp{\textbf{Phase I}: $\tilde g^{(t)}_j$ in \Cref{lin:dp-leaf-sum} in \Cref{alg:sbdt_skeleton}}}
      \HiLi{$\Sigma g \mathrel{+}= \operatorname{sum}_j(\tilde g^{(t)}_j)$\label{alg:earlystop_l6}}\;
      \HiLi{\lIf{$\mathrm{dir} = \texttt{POS}$}{$\Sigma g = \min(\Sigma g,0)$}}\label{alg:earlystop_l7}
      \HiLi{\lElseIf{$\mathrm{dir} = \texttt{NEG}$}{$\Sigma g = \max(\Sigma g,0)$}}\label{alg:earlystop_l8}
      \HiLi{\tcp{\textbf{Phase II}}}
      \HiLi{$\eps_t$ = \RDPAccountant($\dots,\sigma^2_{\text{leaf}},t,\dots$)\label{alg:earlystop_l9}}\;
      \HiLi{$\tau_{\text{CI}}$ = $10^{\eps_t} \cdot \sigma_{\text{CI}}\cdot \tau$\label{alg:earlystop_l10}}\;
      \HiLi{\Switch{$\mathrm{dir}$}{
        \HiLi\Case{\texttt{UNDEF}\label{alg:earlystop_l12}}{
          \HiLi\lIf{$\Sigma g \le -\tau_{\text{CI}}$}{$\mathrm{dir} =\texttt{NEG}$}\label{alg:earlystop_l13}
          \HiLi\lIf{$\Sigma g \ge \tau_{\text{CI}}$}{$\mathrm{dir} =\texttt{POS}$}\label{alg:earlystop_l14}
        }
        \HiLi\Case{\texttt{POS}}{
          \HiLi\lIf{$t \ge 10 \land \Sigma g \le -\tau_{\text{CI}}$}{\KwSty{break\,loop}}\label{alg:earlystop_l16}
        }
        \HiLi\Case{\texttt{NEG}}{
          \HiLi\lIf{$t \ge 10 \land \Sigma g \ge \tau_{\text{CI}}$}{\KwSty{break\,loop}}\label{alg:earlystop_l18}
        }
      }
     }}
    \Return $f$\;\label{alg:earlystop_l19}
\end{algorithm}
\Crefname{algorithm}{Algorithm}{Algorithms}

\Cref{alg:earlystop} leverages our theoretical insight from \Cref{sec:earlystop_theory} for an early tree stopping heuristic that halts as soon as the sum of gradients across the leaves is small. In yellow, we state the new parts compared to the vanilla S-BDT training skeleton of \Cref{alg:sbdt_skeleton}.
Our \textsc{EarlyStoppedSDBT} works in two phases: after training a single tree using S-BDT (cf. \Cref{alg:earlystop_l5}), we (Phase I) build the early stopping criterion $\Sigma g$ solely based on post-processing the sum of noisy gradients across the tree's leaves ($\operatorname{sum}_j(g^{(t)}_j)$) and (Phase II) stop training early if $\Sigma g$ meets the dynamic threshold $\tau_{\text{CI}}$. Since the gradients are noised, we do not halt exactly as in \Cref{eq:stop_crit} ($\operatorname{sum}_j(g^{(t)}_j) = 0$) but when a clear direction shift is detected.

\subsubsection*{Phase I: Build $\Sigma g$ (Post-Processing)}
\Cref{alg:earlystop_l6,alg:earlystop_l7,alg:earlystop_l8} describe $\Sigma g$ as a modified cumulative sum of gradients across the trees:
for instance, we set $\mathrm{dir} = \texttt{POS}$; the gradient sum is positive for the first 20 trees, and it becomes negative starting at tree 21. Then, we take the cumulative sum of all gradients starting from tree 21 until the cumulative sum is larger than or equal to zero again, which repeats this process. This way, we ensure that we only halt if there is a clear change in the sign of the gradient sum, which indicates an extremum in the bias error.
To detect a clear change in the sign of the gradient sum, we define $\mathrm{dir}$ as the sign (positive (POS) or negative (NEG)) of how the gradient sum progresses during early training in \Cref{alg:earlystop_l7,alg:earlystop_l8,alg:earlystop_l12,alg:earlystop_l13,alg:earlystop_l14}.
Thus, we learn whether we should halt on a positive or negative change in the sign of the gradient sum. 
We use the same dynamic cutoff threshold $\tau_{\text{CI}}$ of Phase II to lock in $\mathrm{dir}$. 
We set the gradient sum of a leaf to zero if the noised leaf denominator is smaller than zero (similarly, in this case, S-BDT sets the prediction of this leaf to zero).

\subsubsection*{Phase II: Early Stop If $\Sigma g$ Meets the Dynamic Threshold $\tau_{\text{CI}}$} 
We stop training in \Cref{alg:earlystop_l16,alg:earlystop_l18} based on the cutoff threshold $\tau_{\text{CI}}$, divided into a static ($\tau$, cf. \Cref{alg:earlystop_l2}) and a dynamic part (cf. \Cref{alg:earlystop_l10}).\\
\emph{Static part:} $\tau$ depends on the expected standard deviation of the noise on the per-leaf gradient sum: $\sigma_{\text{leaf}} \cdot g^* \cdot \sqrt{\sfrac{1}{2r_1}}$. As we sum the gradients over $2^d$ leaves, we also sum the noise over them. For $n=2^d$ normally distributed random variables $X_1,\dots,X_n$ with said standard deviation, we know that the sum of these random variables, i.e. $\sum_{i=1}^n X_i$, equals to a random variable with a standard deviation of $\sqrt{\sum_{i=1}^n \sigma_{\text{leaf}}^2 \cdot {g^*}^2 \cdot \sfrac{1}{2r_1}} = \sigma_{\text{leaf}} \cdot g^* \cdot \sqrt{\sfrac{1}{2r_1}} \cdot \sqrt{2^d} \eqqcolon \tau$.\\
\emph{Dynamic part:} Based on $\tau$, we further scale the cutoff with a multiplier of $\sigma_{\text{CI}}$, similar to a confidence interval, to ensure with high certainty that the current change in direction of the gradient sum signal was not due to noise. We additionally find that scaling $\tau$ with $10^{\eps_t}$ provides a dynamic aspect to the early tree stopping, with the goal of not accepting a constant number of trees among all $\eps$ values but more trees for higher $\eps$ values. At higher $\eps$ values, the training uses less noise, so we trust the self-correction of the learning process (i.e., use a higher cutoff $\tau_{\text{CI}}$ that is less sensitive to fluctuations) more than for lower $\eps$ where the noise is high such that one noise sample may destroy all learning process. To determine $\eps$, we call the RDP accountant (cf. \Cref{alg:earlystop_l9}) that we notate in a black-box manner. The RDP accountant tracks the currently consumed privacy budget and, based on the RDP-to-DP conversion \Cref{sec:rdp_conversion}, outputs the $\eps_t$ for the current number of trees $t$.

\subsection{Privacy Implications}
\label{sec:privacy_earlystopping}

Our early stopping is adaptive to the learning task by only using S-BDT's DP approximations of gradient sums ($\Sigma g$). This constitutes post-processing, which incurs no extra privacy leakage (cf. \Cref{lem:postprocessing}). A mechanism also remains RDP if it adaptively uses the output of a previous one (so-called adaptive sequential composition) \cite[Proposition 1]{DBLP:journals/corr/Mironov17}.

\subsubsection*{Early Stopping and Privacy Filters} 
If we stop S-BDT training early, we do not leak privacy for trees that we did not train.
However, simply reducing the privacy budget $\rho(\alpha)$ by the number of skipped trees does not necessarily imply DP with this $\rho(\alpha)$ since DP is an a priori guarantee before the mechanism is executed. Yet, it is inherent to early stopping that we do not know beforehand when we will stop. Thus, in early stopping, we can only guarantee this property before training the next tree, since only then do we know that we did not stop after the current tree. In an extreme case, an attacker could choose a neighboring dataset pair such that early stopping never works, and we always train for $T$ rounds, which still preserves DP but with the original privacy budget.

So-called \emph{Privacy filters} \cite{rogers2016privacy,feldman2020individual,lecuyer2021practical} offer a solution to this a-posteriori leakage saving. S-BDT already uses an RDP privacy filter \cite{feldman2020individual} for continuous learning.
In this privacy filter, we can run arbitrarily many adaptive RDP mechanisms, S-BDT among them, and only halt if a previously defined privacy budget $\rho(\alpha)$ is exceeded. For that, we use the $\rho(\alpha)$ associated with the maximal number of trees as this budget, but can use the $\rho(\alpha)$ saved by stopping early for other RDP mechanisms on the same dataset. For compatibility, we only report the $\rho(\alpha)$ spent on S-BDT in our experiments.

\begin{figure*}
    \centering
    \begin{minipage}[t]{.325\textwidth}
        \begin{subfigure}[h]{\textwidth}
            \resizebox{\textwidth}{!}{\input{images/insurance_utility_privacy_6000.pgf}\unskip}
            \caption{Test (\emph{Insurance}), $\downarrow$ is better}
        \end{subfigure}
        \\[0.5em]
        \begin{subfigure}[h]{\textwidth}
            \resizebox{\textwidth}{!}{\input{images/credit1_utility_privacy_6000.pgf}\unskip}
            \caption{Test (\emph{Credit}), $\uparrow$ is better}
        \end{subfigure}
        \captionof{figure}{\textbf{Main Result}: Privacy-utility tradeoff on test data comparing our early stopping (orange) against the baseline of purely statically chosen hyperparameters (blue).
        Top: Regression; Bottom: Classification.
        }
        \label{fig:earlystop_1}
    \end{minipage}
    \hfill
    \begin{minipage}[t]{.65\textwidth}
        \begin{subfigure}[h]{0.495\textwidth}
            \resizebox{\textwidth}{!}{\input{images/abalone_utility_privacy_6000.pgf}\unskip}
            \caption{Comparison (\emph{Abalone}), $\downarrow$ is better}
        \end{subfigure}
        \hfill
        \begin{subfigure}[h]{0.495\textwidth}
            \resizebox{0.99\textwidth}{!}{\phantom{~}}
        \end{subfigure}
        \\[0.5em]
        \begin{subfigure}[h]{0.495\textwidth}
            \resizebox{\textwidth}{!}{\input{images/adult_utility_privacy_6000.pgf}\unskip}
            \caption{Comparison (\emph{Adult}), $\uparrow$ is better}
        \end{subfigure}
        \hfill
        \begin{subfigure}[h]{0.495\textwidth}
            \resizebox{\textwidth}{!}{\input{images/spambase_utility_privacy_6000.pgf}\unskip}
            \caption{Comparison (\emph{Spambase}), $\uparrow$ is better}
        \end{subfigure}
        \captionof{figure}{
            \textbf{Comparison to HP-tuned S-BDT}: Privacy-utility tradeoff on comparison data comparing our early stopping (orange) against the baselines of purely statically chosen hyperparameters (blue) and tuned yet privacy-leaky hyperparameters (green).
            Top: Regression; Bottom: Classification.
        }
        \label{fig:earlystop_2}
    \end{minipage}
    \label{fig:earlystop}
\end{figure*}

\section{Evaluation}
\label{sec:evaluation}

We test MammothDP's full system with its key components---constant-time implementation of S-BDT, improved rollback protection, and early tree stopping---on two TEEs: a \emph{TDX}-enabled \emph{Intel Xeon Gold 6526Y Emerald Rapids} and an \emph{SEV}-enabled \emph{AMD EPYC 7763} processor.
We provide an Ubuntu 24.10 image with a compiled MammothDP binary\footnote{\url{https://github.com/kirschte/MammothDP/tree/main/tee-ready-image}} and MammothDP's code\footnote{\url{https://github.com/kirschte/MammothDP}}.

\subsubsection*{Structure}
We extensively evaluate \emph{MammothDP} with regard to constant-time verification (cf. \Cref{sec:microwalkevaluation}), early stopping (cf. \Cref{sec:eval_early_stopping}), and runtime overhead (cf. \Cref{sec:runtime-overhead}).

\subsection{Constant-Time Verification}
\label{sec:microwalkevaluation}

Constant-time code does not necessarily guarantee security against timing attackers. For example, subtle leakages may be missed~\cite{DBLP:conf/ccs/SieckBW021,DBLP:conf/uss/WeiserSBS20}, or compiler optimizations may reintroduce vulnerabilities~\cite{DBLP:journals/tissec/DanielBR23,DBLP:conf/eurosp/SimonCA18}.
Thus, we must verify compiled binaries. 
We use Microwalk~\cite{microwalk,DBLP:conf/ccs/WichelmannSP022}, a dynamic timing side-channel analysis framework.
For Microwalk's analysis, see \Cref{sec:additional-preliminaries-microwalk}.

For the analysis, we divided data-dependent parts of our implementation into independent components.
For each component, we generated 100 random test inputs. These are loaded by thin wrapper functions, which transform the test inputs into the correct format (while maintaining uniformity) and then call the target component. 
We embed the wrapper functions into Microwalk's new analysis template~\cite{DBLP:conf/cans/FayolleWKREM24}, which automates communication with Microwalk and loading and executing the test inputs.
The resulting analysis workflow is fully automatic, which makes re-verifying the full implementation after changes straightforward.

Microwalk verifies all components, without reporting leakages. 
The analysis covered all executed S-BDT code, and increasing the number of test cases does not raise coverage. We conclude that our implementation exhibits no data-dependent control flow or memory accesses, making it robust against system-level attackers.

\subsection{Early Stopping (Privacy-Utility Tradeoff)}
\label{sec:eval_early_stopping}

We test the privacy-utility tradeoff on five datasets (cf. \Cref{tbl:datasets}) that we describe in \Cref{sec:dataset_details} in more detail.

\begin{table}[!ht]
    \centering
    \caption{Sensitive Datasets.}
    \label{tbl:datasets}
    \begin{tabular}{lcrr}
        \toprule
        Name & Type & Size & \#Attributes \\
        \midrule
        Insurance~\cite{insurance} & Regression & 1,300 & 6 \\
        Abalone~\cite{misc_abalone_1} & Regression & 4,200 & 8 \\
        Credit~\cite{GiveMeSomeCredit} & Classification & 120,300 & 10 \\
        Adult~\cite{misc_adult_2} & Classification & 48,800 & 14 \\
        Spambase~\cite{misc_spambase_3} & Classification & 4,600 & 57 \\
       \bottomrule
  \end{tabular}
\end{table}

\subsubsection{How does the privacy-utility tradeoff of early stopped S-BDT compare?}

In \Cref{fig:earlystop_1,fig:earlystop_2}, we compare S-BDT with leakage-free hyperparameters in the vanilla and early stopping (cf. \Cref{alg:earlystop} on $\sigma_{\text{CI}}=3$) variant, and, as a reference, we include S-BDT with privacy-leaky hyperparameters \cite[Figure 3]{sbdt}. For compatibility, we only report the $\rho(\alpha)$---converted to $(\eps,\delta)$-DP---spent on S-BDT (cf. \Cref{sec:privacy_earlystopping} for the Rényi Filter discussion). We also bucket similar actually consumed privacy budgets that vary due to early stopping into bins for better plot readability.

\boldparagraph{Experimental Setup.}
We set $\delta = 5 \cdot 10^{-8}$.
We evaluate early stopping with 20 runs of 5-fold cross-validation, except for the \emph{Insurance} (100 runs) and \emph{Abalone} (50 runs) datasets.
For both S-BDT variants with leakage-free hyperparameters, we chose a static set of hyperparameters based on the compromise of the best-performing hyperparameters in HP-tuned S-BDT on \emph{Abalone}, \emph{Adult}, and \emph{Spambase}. \Cref{sec:static_hyperparameters} lists them.

\boldparagraph{Comparison vs. Test Datasets.} Our three comparison datasets (\emph{Abalone}, \emph{Adult}, and \emph{Spambase}) are the same as in S-BDT \cite{sbdt} and serve to determine the static hyperparameters.
For comparability, we reuse S-BDT's attribute and regression value ranges, which are listed in \Cref{sec:static_hyperparameters}.
To illustrate the performance gain in the leakage-free hyperparameter setting of early-stopped and vanilla S-BDT, we chose the two test datasets (\emph{Insurance} and \emph{Credit}) that were not used to determine the static set of reasonable hyperparameters, and attribute pre-processing has been limited to taking a numerical attribute range and a regression value range on abstract metadata.

\boldparagraph{Results.}
In the leakage-free hyperparameter setting, early-\linebreak[4]stopped S-BDT significantly outperforms non-HP-tuned S-BDT on both regression datasets, regardless of whether test or comparison datasets are used. For classification, early stopping only outperforms S-BDT in the lower $\eps$ regime. For all comparison datasets, we reach a lower but comparable utility when comparing early-stopped S-BDT to the reference S-BDT with privacy-leaky hyperparameters.

\boldparagraph{Ablation on (maximal) Number of Trees $T$.}
We show in \Cref{fig:exp_60_600_600_trees_1,fig:exp_60_600_600_trees_2} how other initial guesses of the current (maximal) number of trees $T=6{,}000$ impact the privacy-utility tradeoff.

\subsubsection{Is early stopping actually adaptive?}
\label{sec:eval_adaptivity_earlystop}
In \Cref{fig:exp_real_nb_trees_1,fig:exp_real_nb_trees_2}, we visualize the actual number of chosen trees for all three variants. These show for the regression datasets, that early stopping does indeed select the number of trees adaptively and that, as designed, early-stopped S-BDT selects more trees, the larger $\eps$ grows. A similar, although less distinct, pattern is also present in HP-tuned S-BDT. For classification, the adaptive nature of early stopped S-BDT is visible for \emph{Spambase} but not for \emph{Credit} and \emph{Adult}.

\subsubsection{Is it justified to assume that A-I) 15 hyperparameters can be fixed statically across $\varepsilon$ values and datasets and A-II) the number of trees cannot?} We illustrates both in \Cref{fig:exp_60_600_600_trees_2} on comparison data (where static hyperparameters can be correlated with the aggregate of all optimal ones, cf. \Cref{sec:eval_early_stopping}): (A-I) the gap between HP-tuned (green, tuned per $\varepsilon$ and dataset) and non-HP-tuned (blue, static hyperparameters) is small, (A-II) but only for the right number of trees which varies strongly per $(\varepsilon, \text{dataset})$: (high $\varepsilon$, adult) 6000 trees, (low $\varepsilon$, adult) 600 trees, (high $\varepsilon$, abalone) 60 trees.

\subsection{Runtime overhead}
\label{sec:runtime-overhead}

\subsubsection{What is \emph{MammothDP}'s runtime overhead?}

\begin{figure}[!t]
  \centering
  \begin{subfigure}[h]{\linewidth}
    \resizebox{\textwidth}{!}{\includegraphics{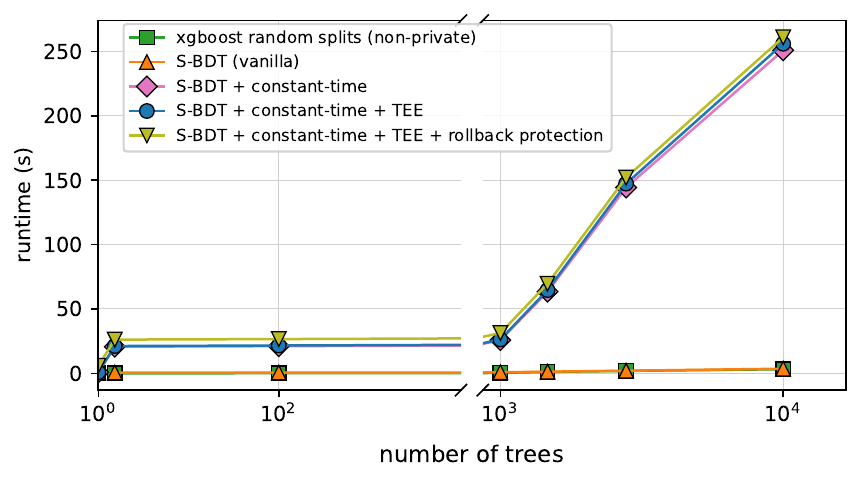}\unskip}
      \caption{Training runtime on \emph{Abalone} vs. number of trees.}
      \label{fig:runtime-vs-number-of-trees}
  \end{subfigure}
    \\[0.5em]
  \begin{subfigure}[h]{\linewidth}
    \resizebox{\textwidth}{!}{\includegraphics{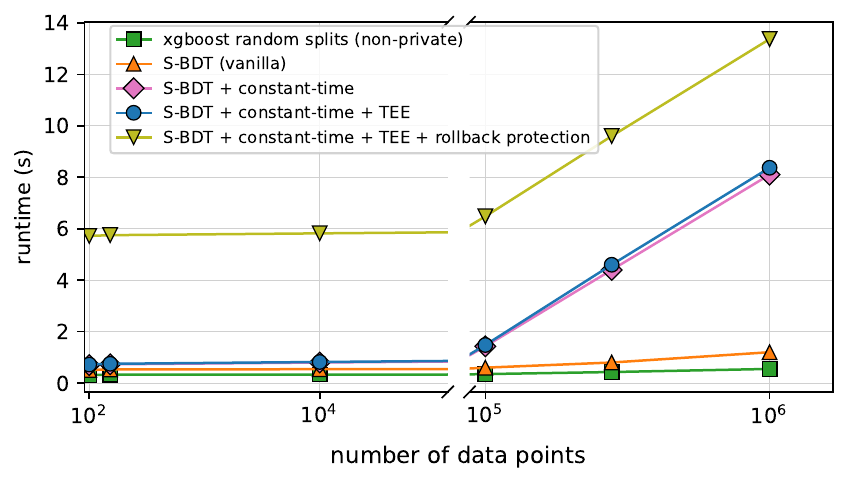}\unskip}
      \caption{Training runtime of one tree vs. number of data points.}
      \label{fig:runtime-vs-number-of-datapoints}
  \end{subfigure}
    \hfill
  \caption{\textbf{Runtime} in seconds of MammothDP's components. MammothDP consists of a constant-time implementation of S-BDT with early tree stopping, a TEE port, and rollback protection.}
  \label{fig:runtime-vs}
\end{figure}

In \Cref{fig:runtime-vs}, we compare MammothDP's runtime overhead to vanilla S-BDT and non-private XGBoost \cite{chen2016xgboost}, and pin down the overhead of the constant-time implementation, the TEE, and the rollback protection.

Our measurements highlight one key effect: the most runtime overhead is due to the constant-time implementation, which scales linearly in both dataset size and number of trees. The overhead of the constant-time implementation is due to fixed-time instructions taking more clock cycles to run \cite[Figure 14]{libfixedtimefixedpoint}, and linear scans over bitmasks of the entire dataset instead of just the current subsample.

For rollback protection alone, we measure $0{.}5$ seconds for initialization, $5{.}0$ seconds for one training call, and $5{.}0$ seconds for recovery of benign execution crashes. Most runtime is used for authentication and nonce verification. Sealing and unsealing only take 0.001 seconds each. Using a TEE incurs $0{.}02 - 5$ seconds overhead.

\subsubsection{How does \emph{MammothDP}'s runtime overhead compare to other solutions?}

Our constant-time S-BDT implementation incurs factor $1{.}6 - 69{.}4$ overhead over vanilla S-BDT. Oblivious Secure XGBoost \cite{xgboostontrustedhw}, a constant-time, secure and collaborative version of XGBoost \cite{chen2016xgboost}, incurs a factor of $16{.7} - 178{.}2$ overhead over vanilla Secure XGBoost. The specific overhead depends on a hyperparameter that must be tuned. Furthermore, our work, unlike Oblivious Secure XGBoost, also defends against attacks based on instruction runtime, which gives us an additional overhead. 

TEE and rollback protection of MammothDP incur a factor of $2{.}8 - 8{.}3$ overhead over vanilla S-BDT. ElephantDP \cite{elephantdp}, a state-of-the-art rollback protection for DP algorithms, incurs a factor of $1{.}1 - 3{.}9$ overhead over vanilla DP approximated average, variance, and Pearson correlation coefficient, and private multiplicative weights. 

Our TEE setup with rollback protection takes $5{.}3 - 5{.}7$ seconds for training one tree with vanilla S-BDT. MPC-based GBDT solutions \cite{SoCuBaChShRu_25:Guard-GBDT,JiMeDaLi_24:SiGBDT} take $29-865$ seconds to train one tree.

\section{Conclusion}
\label{sec:conclusion}

We introduced \emph{MammothDP}, a significant milestone towards a deployable DP system for data spaces that protects the processing of data and the release of studies alike. By combining DP, TEEs, and constant-time programming, MammothDP addresses privacy and security requirements relevant to the European Health Data Space.
MammothDP is not a one-size-fits-all TEE-based DP tree ensemble learner but shows a design that can reasonably trade off partially contradicting goals of a data space: low computational overhead, low communication overhead, strong privacy guarantees, limited utility loss, no TEE rollbacks, no required Rényi filter, and timing side-channel protection.
In detail, MammothDP utilizes two novel components: an improved rollback protection and a post-processing-based early stopping rule that allows adaptive hyperparameter tuning without extra privacy budget and thus attains an improved privacy-utility tradeoff for released studies.

\bibliographystyle{IEEEtran}
\bibliography{bibliography}

\appendices
\crefalias{section}{appendix}
\crefalias{subsection}{subappendix}
\crefalias{subsubsection}{subsubappendix}

\section*{Appendix Overview}

\startcontents
\printcontents{}{1}{}

\begin{figure*}[!ht]
    \centering
    \begin{minipage}[t]{0.325\textwidth}        
        \begin{subfigure}[h]{\textwidth}
            \resizebox{\textwidth}{!}{\input{images/insurance_real_nb_trees_6000.pgf}\unskip}
            \caption{Test (\emph{Insurance})}
        \end{subfigure}
        \\[0.5em]
        \begin{subfigure}[h]{\textwidth}
            \resizebox{\textwidth}{!}{\input{images/credit1_real_nb_trees_6000.pgf}\unskip}
            \caption{Test (\emph{Credit})}
        \end{subfigure}
        \captionof{figure}{
            \textbf{Actual number of trees} on test data for our early-stopped S-BDT (orange) and, as a reference, the S-BDT baseline of purely statically chosen hyperparameters (blue).
            Top: Regression; Bottom: Classification.
        }
        \label{fig:exp_real_nb_trees_1}
    \end{minipage}
    \hfill
    \begin{minipage}[t]{0.65\textwidth}
        \begin{subfigure}[h]{0.495\textwidth}
            \resizebox{\textwidth}{!}{\input{images/abalone_real_nb_trees_6000.pgf}\unskip}
            \caption{Comparison (\emph{Abalone})}
        \end{subfigure}
        \hfill
        \begin{subfigure}[h]{0.495\textwidth}
            \resizebox{0.99\textwidth}{!}{\phantom{~}}
        \end{subfigure}
        \\[0.5em]
        \begin{subfigure}[h]{0.495\textwidth}
            \resizebox{\textwidth}{!}{\input{images/adult_real_nb_trees_6000.pgf}\unskip}
            \caption{Comparison (\emph{Adult})}
        \end{subfigure}
        \hfill
        \begin{subfigure}[h]{0.495\textwidth}
            \resizebox{\textwidth}{!}{\input{images/spambase_real_nb_trees_6000.pgf}\unskip}
            \caption{Comparison (\emph{Spambase})}
        \end{subfigure}
        \captionof{figure}{
            \textbf{Actual number of trees} on comparison data for our early-stopped S-BDT (orange) and, as a reference, the S-BDT baselines of purely statically chosen hyperparameters (blue) and tuned yet privacy-leaky hyperparameters (green).
            Top: Regression; Bottom: Classification.
            Note that the zigzags in HP-tuned S-BDT might be an artifact of the HP-search where 1) the number of trees is discretized and 2) a different number of trees might only be infinitesimally worse than the reported one.}
        \label{fig:exp_real_nb_trees_2}
    \end{minipage}
\end{figure*}

\section{Non-TEE-based Related Work}
\label{sec:nontee_relatedwork}

\subsubsection*{Secure DP-ML processing without a TEE but an MPC protocol}
Secure multiparty computation protocols (MPC, a.k.a. secure computation) would technically provide an alternative avenue towards confidential, trustworthy, and accountable execution of a DP mechanism. Admittedly, MPC-based approaches would be free from vendor-specific vulnerabilities and having to trust the vendors. Yet, some of the computation parties would instead need to be trusted. To prevent malicious noise generation, the MPC implementation would need to include secure noise generation. The rollback protection would be ensured by trusting some of the computation parties and by requiring that the MPC protocol is secure against active attackers. As our early stopping mechanism requires a private threshold check, relying on secure aggregation alone (as in~\cite{Maddock_2022,sbdt}) would not be sufficient. Hence, the entire computation of all trees (cf. \Cref{fig:exp_real_nb_trees_1,fig:exp_real_nb_trees_2} for the number of trees) would have to be done in MPC. As a result, the computational costs would be orders of magnitude higher (even when including the overhead due to the constant-time implementation and rollback protection) and it would incur a very large communication cost overhead, with existing MPC protocols from the literature~\cite{JiMeDaLi_24:SiGBDT,SoCuBaChShRu_25:Guard-GBDT}.

\subsubsection*{Secure DP-ML processing without a TEE but an accountable computation protocol}
Another line of related work is that of accountable computation~\cite{BaDrHaUn_09:csar,BaFiMo_13:ppac}. Those results solely rely on cryptographic assumptions and a secure logging mechanism. Yet, to the best of our knowledge, existing work does not enable accountable computation of DP algorithms. Moreover, these accountable computation approaches do not tackle the topic of rollback protection, which is essential for satisfying differential privacy.

\section{Detailed Background}

\subsection{Microwalk}
\label{sec:additional-preliminaries-microwalk}

There has been much research in automated constant-time verification, and many tools have been proposed~\cite{DBLP:conf/ccs/GeimerVRDBM23}. One of these tools is Microwalk~\cite{microwalk,DBLP:conf/ccs/WichelmannSP022}, a dynamic timing side-channel analysis framework which has, since its initial proposal, matured to a stable and practically usable tool. 

Microwalk uses a dynamic approach, i.e., it analyzes a program for a concrete set of random inputs, which are carefully crafted to achieve high implementation coverage. For every such input, Microwalk collects an execution trace with all taken branches and accessed memory addresses, which resembles an overapproximation of the information a powerful timing side-channel attacker can acquire. The subsequent timing side-channel leakage analysis follows a simple idea: If all these traces are \emph{identical}, the random inputs led to the same execution path, i.e., they are indistinguishable for a timing side-channel attacker. On the other hand, if there are \emph{differences} between these traces, the attacker can distinguish the inputs and thus learn (partial) information. Microwalk compares execution traces and computes a leakage score, indicating how much information a timing side-channel attacker would get in the worst case. To aid fixing the identified vulnerabilities, Microwalk outputs information about the source code location and witness inputs that trigger the vulnerability.

\subsection{FACCT Discrete Gaussian Sampling}
\label{sec:additional-preliminaries-facct}

For a discrete Gaussian $D_\sigma$ with $\sigma = k\cdot \sigma_0$ and $\sigma_0 = \sqrt{1/(2\ln{}2)}$, FACCT uses the binary sampling method of the BLISS signature scheme \cite{binarysamplingmethod}: first call the base sampler to generate a sample $x \sim \mathcal{D}_{\sigma_0}^+$ from a positive discrete Gaussian with standard deviation $\sigma_0$. The base sampler uses an adaptation of the inversion method from Bos et al. \cite{ct_base_sampler}. The inversion method samples $x \sim \text{pdf}_f$ from a probability density function $\text{pdf}_f$ by sampling $u \sim \mathcal{U}([0,1])$ and then computing $F^{-1}_{f}(x)$ with the inverse CDF $F^{-1}_f$. This approach can be implemented in constant-time using a full-table access CDT sampler.

Now, generate $y \sim \mathcal{U}(\{0,1,...,k-1\})$, set $z \coloneqq kx + y$, $t \coloneqq y(y + 2kx)$ and perform rejection sampling on $z$ with rejection rate $p = \exp{(-y(y + 2kx)/(2\sigma^2))}$ using a Bernoulli sampler. The Bernoulli sampler needs to compute $p$ first to generate a sample. Zhao et al. \cite{facct} propose to compute a polynom $P$ as an approximation of $2^x$ for $0\leq x < 1$ which is then further used to compute $p$ (cf. \cite[Figure 8]{facct}). Zhao et al. find a polynom $P$ that has maximal relative error $\max{}_x|\frac{P(x) - 2^x}{2^x}| < 2^{-45}$. They evalute $P$ which has degree $n$ and coefficients $(a_i)_{i=1}^{n}$ by using Horner's rule ($H(x)$) with $\delta_\text{FP}$-bit precision, yielding absolute error $|P(x) - H(x)| \leq \gamma_{2n} \cdot \sum_{i=0}^{n} |a_i| \cdot |x|^i$, where $\gamma_{2n} \approx 2n \cdot 2^{-\delta_\text{FP}}$. Evaluating the polynom is possible in constant-time because the degree and precision is fixed.

The probability of outputting an integer $z = kx + y$ is now proportional to $\exp{}(-z^2/(2\sigma^2))$ (\cite[Theorem 3]{facct}).

\subsection{RDP-to-DP conversion}
\label{sec:rdp_conversion}
With \Cref{cor:rdp_to_dp}, we can convert RDP to DP. For a tight conversion, we refer to \cite[Theorem 2]{sommer2019privacy} which requires access to all $\alpha$.

\begin{corollary}[RDP to DP {\cite[Thm. 21]{balle2019hypothesis}}, {\cite[Prop. 12]{canonne2020discrete}}]\label{cor:rdp_to_dp}
    For any $\delta \in [0,1]$, if a mechanism is $(\alpha,\rho(\alpha))$-RDP, then it is $(\eps,\delta)$-DP with
    $\eps = \rho(\alpha) + \log(\frac{\alpha - 1}{\alpha}) - \frac{\log(\delta) + \log(\alpha)}{\alpha - 1}$.
\end{corollary}

\section{Extended Evaluation}
\label{sec:extended-evaluation}

\subsection{Dataset Description (cf. Table~\protect\ref{tbl:datasets})}
\label{sec:dataset_details}

\begin{itemize}
    \item \emph{Insurance} predicts the medical insurance premium of the beneficiary based on attributes like age, sex, and children.
    \item \emph{Abalone} predicts the age of an Abalone based on attributes like sex, length, and diameter.
    \item \emph{Credit} predicts whether a borrower had a delinquent payment based on attributes like a borrower's monthly income. Here, we removed N/A entries and unlabelled data points.
    \item \emph{Adult} predicts whether a person earns over 50,000\,\$ per year based on attributes like age, sex, and occupation.
    \item \emph{Spambase} predicts whether mail is spam based on attributes like the frequencies of words and symbols in a mail.
\end{itemize}

\subsection{Static Hyperparameters}
\label{sec:static_hyperparameters}
For both S-BDT variants with leakage-free hyperparameters, we chose the following static set of hyperparameters based on the compromise of the best-performing hyperparameters in HP-tuned S-BDT on the \emph{Abalone}, \emph{Adult}, and \emph{Spambase} datasets:
\begin{itemize}
    \item learning rate $\eta = 0.1$,
    \item maximal number of trees $T = 6{,}000$,
    \item depth of trees $d=2$,
    \item adaptive leaf noise ratio $r_1=0.4$,
    \item gradient clipping bound $g^*=0.2$,
    \item Hessian clipping bound $h^*=0.2$,
    \item subsampling rate $\gamma=0.2$,
    \item leaf regularization $\lambda=15$,
    \item leaf regularization mode \texttt{ADD},
    \item \emph{enabled} cyclical feature interaction,
    \item \emph{enabled} ignoring the split constraints,
    \item \emph{enabled} random splits from candidates,
    \item \emph{disabled} split refinement,
    \item \emph{enabled} initial score solely for regression datasets, and
    \item initial score clipping bound $m^*=0.5$.
\end{itemize}
For comparability on the three comparison datasets, we reuse S-BDT's attribute and regression value ranges: for \emph{Abalone}, the regression value range is the actual min-max values and the attribute range is $0\dots0.5$; for \emph{Adult} $0\dots100$; and for \emph{Spambase} $0\dots1$.

\begin{figure*}[!ht]
    \centering
    \begin{minipage}{\textwidth}
        \begin{subfigure}[h]{0.299\textwidth}
            \resizebox{\textwidth}{!}{\input{images/insurance_utility_privacy_60.pgf}\unskip}
            \caption{Test (\emph{Insurance}), $T=60$}
        \end{subfigure}
        \hfill
        \begin{subfigure}[h]{0.299\textwidth}
            \resizebox{\textwidth}{!}{\input{images/insurance_utility_privacy_600.pgf}\unskip}
            \caption{Test (\emph{Insurance}), $T=600$}
        \end{subfigure}
        \hfill
        \begin{subfigure}[h]{0.299\textwidth}
            \resizebox{\textwidth}{!}{\input{images/insurance_utility_privacy_6000.pgf}\unskip}
            \caption{Test (\emph{Insurance}), $T=6{,}000$}
        \end{subfigure}
        \\[0.5em]
        \begin{subfigure}[h]{0.299\textwidth}
            \resizebox{\textwidth}{!}{\input{images/credit1_utility_privacy_60.pgf}\unskip}
            \caption{Test (\emph{Credit}), $T=60$}
        \end{subfigure}
        \hfill
        \begin{subfigure}[h]{0.299\textwidth}
            \resizebox{\textwidth}{!}{\input{images/credit1_utility_privacy_600.pgf}\unskip}
            \caption{Test (\emph{Credit}), $T=600$}
        \end{subfigure}
        \hfill
        \begin{subfigure}[h]{0.299\textwidth}
            \resizebox{\textwidth}{!}{\input{images/credit1_utility_privacy_6000.pgf}\unskip}
            \caption{Test (\emph{Credit}), $T=6{,}000$}
        \end{subfigure}
        \captionof{figure}{S-BDT on 2 test datasets: RMSE/AUC vs. (maximal) number of trees $T$. Top: regression; bottom: classification.}
        \label{fig:exp_60_600_600_trees_1}
    \end{minipage}
    \\[0.5em]
    \begin{minipage}{\textwidth}
        \begin{subfigure}[h]{0.299\textwidth}
            \resizebox{\textwidth}{!}{\input{images/abalone_utility_privacy_60.pgf}\unskip}
            \caption{Comparison (\emph{Abalone}), $T=60$}
        \end{subfigure}
        \hfill
        \begin{subfigure}[h]{0.299\textwidth}
            \resizebox{\textwidth}{!}{\input{images/abalone_utility_privacy_600.pgf}\unskip}
            \caption{Comparison (\emph{Abalone}), $T=600$}
        \end{subfigure}
        \hfill
        \begin{subfigure}[h]{0.299\textwidth}
            \resizebox{\textwidth}{!}{\input{images/abalone_utility_privacy_6000.pgf}\unskip}
            \caption{Comparison (\emph{Abalone}), $T=6{,}000$}
        \end{subfigure}
        \\[0.5em]
        \begin{subfigure}[h]{0.299\textwidth}
            \resizebox{\textwidth}{!}{\input{images/adult_utility_privacy_60.pgf}\unskip}
            \caption{Comparison (\emph{Adult}), $T=60$}
        \end{subfigure}
        \hfill
        \begin{subfigure}[h]{0.299\textwidth}
            \resizebox{\textwidth}{!}{\input{images/adult_utility_privacy_600.pgf}\unskip}
            \caption{Comparison (\emph{Adult}), $T=600$}
        \end{subfigure}
        \hfill
        \begin{subfigure}[h]{0.299\textwidth}
            \resizebox{\textwidth}{!}{\input{images/adult_utility_privacy_6000.pgf}\unskip}
            \caption{Comparison (\emph{Adult}), $T=6{,}000$}
        \end{subfigure}
        \\[0.5em]
        \begin{subfigure}[h]{0.299\textwidth}
            \resizebox{\textwidth}{!}{\input{images/spambase_utility_privacy_60.pgf}\unskip}
            \caption{Comparison (\emph{Spambase}), $T=60$}
        \end{subfigure}
        \hfill
        \begin{subfigure}[h]{0.299\textwidth}
            \resizebox{\textwidth}{!}{\input{images/spambase_utility_privacy_600.pgf}\unskip}
            \caption{Comparison (\emph{Spambase}), $T=600$}
        \end{subfigure}
        \hfill
        \begin{subfigure}[h]{0.299\textwidth}
            \resizebox{\textwidth}{!}{\input{images/spambase_utility_privacy_6000.pgf}\unskip}
            \caption{Comparison (\emph{Spambase}), $T=6{,}000$}
        \end{subfigure}
        \captionof{figure}{S-BDT on 3 comparison datasets: RMSE/AUC vs. (maximal) number of trees $T$. Top: regression; middle/bottom: classification.}
        \label{fig:exp_60_600_600_trees_2}
    \end{minipage}
\end{figure*}

\subsection{Adaptive Number of Trees}
In \Cref{fig:exp_real_nb_trees_1,fig:exp_real_nb_trees_2} we show how the actual number of trees of all three S-BDT variants changes throughout the training (cf. \Cref{sec:eval_early_stopping} for the interpretation).

\subsection{Ablation on (Maximal) Number of Trees $T$}
\label{sec:ablation}

We show in \Cref{fig:exp_60_600_600_trees_1,fig:exp_60_600_600_trees_2} how the privacy-utility tradeoff of early-stopped and non-HP-tuned S-BDT changes when we use a more or less sophisticated initial guess of the (maximal) number of trees $T$. 
If we select a small number of trees $T=60$, both variants of S-BDT improve in the lower $\eps$ regime but worsen in the higher $\eps$ regime. If we select a high number of trees $T=6{,}000$, then non-HP-tuned S-BDT worsens in the lower $\eps$ regime but improves in the higher $\eps$ regime.
In comparison, early-stopped S-BDT has an RMSE/AUC roughly on par with non-HP-tuned S-BDT for a small $T=60$ except for \emph{Abalone}, where the early-stopped S-BDT is better. However, early-stopped S-BDT has a significantly better RMSE/AUC for the majority of (dataset-$\eps$)-combinations with a growing number of trees starting from $T=600$ for regression and $T=6{,}000$ for classification. This advantage of early-stopped S-BDT is especially visible in the lower $\eps$ regime.

We chose $T=6{,}000$ trees as a default for our early-stopped S-BDT experiments, since our experiments show that non-HP-tuned S-BDT sometimes performs better on $T=6{,}000$ trees.
As seen in \Cref{fig:exp_60_600_600_trees_1,fig:exp_60_600_600_trees_2}) when comparing the blue lines of the center with the one of the right column and whenever $\eps$-values are in range, starting from roughly $\eps=4.3$ (\emph{Insurance}), $\eps=1.1$ (\emph{Credit}), $\eps=0.11$ (\emph{Adult}), and $\eps=0.5$ (\emph{Spambase}) the RMSE/AUC of non-HP-tuned S-BDT for $T=600$ trees on par or worse with the one with $T=6{,}000$ trees.
At $\eps=10$ and $T=6{,}000$, non-HP-tuned S-BDT has a decisively better RMSE/AUC then $T=600$ trees: $6600$ vs $7200$ (\emph{Insurance}), $84.4\,\%$ vs $85.2\,\%$ (\emph{Credit}), $88.4\,\%$ vs $90.0\,\%$ (\emph{Adult}), and $95.2\,\%$ vs $97.1\,\%$ (\emph{Spambase}). Only \emph{Abalone} falls short in the RMSE: $2.44$ ($T=600$) vs $2.48$ ($T=6{,}000$).

\section{Rollback Protection Details}
\label{sec:rollback-appendix}

\subsection{System and Adversary Model}

We consider the following entities shown in Figure~\ref{fig:rollback-overview} of Section~\ref{sec:rollback}. The DP system is hosted on an untrusted \emph{server}. The server's administrator may be malicious, or the server's privileged software (OS, hypervisor) may be compromised. The server has a trusted execution environment (\emph{TEE}) that provides secure execution, but may leak timing side-channel traces to the adversary. The TEE supports remote attestation that allows external entities to verify the code of the TEE application and securely send data to an attested TEE application. Intel's SGX or TDX are examples of such TEE architectures.

\emph{Data owners} attest the TEE application before sending sensitive training data to it. Data owners can be either honest or malicious. Honest data owners follow the protocol to protect their data. Malicious data owners may deviate from the protocol in an attempt to violate the privacy of other data owners. We assume that at least one of the data owners is honest. If all data owners are malicious, no solution can protect the privacy of their data.

The server has a \emph{disk} to which the TEE application can save its persistent state using the TEE's sealing functionality. We assume that sealing supports data confidentiality and authenticity, but not data freshness, which is the case in current commodity TEEs. Finally, we assume a state continuity module (\emph{SCM}) which could be a local security chip, such as a TPM, or a distributed state-keeping system~\cite{angel2023nimble,niu2022narrator} that is accessed over a network connection. 

\subsubsection*{SCM interface} The SCM provides the following interface that can be used by the TEE and other entities. For example, in TPMs, this interface is available through the password-based authorization features that control access to non-volatile memory (NVRAM) indices. 

\begin{itemize}
    \item \textsf{SCM.Authorize(pwd)}: This operation prepares the SCM for subsequent use by any entity that knows the password \textsf{pwd}. We assume that the TEE can call this operation without leaking the \textsf{pwd} to the adversary.
    
    \item \textsf{\textsf{ok/fail} $\leftarrow$ SCM.Write(data, pwd)}: This operation writes \textsf{data} to the SCM. If the SCM was previously authorized using \textsf{pwd}, the operation returns \textsf{ok}. Otherwise, the operation returns \textsf{fail}.
    
    \item \textsf{data/fail} $\leftarrow$ \textsf{SCM.Read(pwd)}: This operation returns the current content of the SCM,  assuming that the SCM was previously authorized using \textsf{pwd}. Otherwise, this operation returns \textsf{fail}.
\end{itemize}
        
\subsubsection*{Sealing interface} We assume a sealing mechanism with the following interface. We note that in TEEs like Intel's SGX, the architecture provides sealing and unsealing calls for TEE applications. In other TEE architectures, such as Intel's TDX, the architecture only derives the sealing key for the protected VM, which must then implement encryption and authentication of data itself.

\begin{itemize}
    \item \textsf{TEE.Seal(ID, data)}: This operation stores data identified by \textsf{ID} to disk using authenticated encryption based on a key provided by the TEE architecture.

    \item \textsf{data/fail} $\leftarrow$ \textsf{TEE.Unseal(ID)}: This operation reads, decrypts, and authenticates sealed data identified by \textsf{ID} from disk. If authentication succeeds, the operation returns the decrypted \textsf{data}. Otherwise, this operation returns \textsf{fail}.
\end{itemize}

\subsubsection*{Adversary model} We consider a privileged adversary that controls the untrusted server. The adversary can start and stop the TEE execution freely. During the TEE execution, the adversary can collect timing side-channel traces. The adversary cannot break the TEE's execution integrity or directly access any secret data processed inside the TEE. Neither can the adversary break the remote attestation.

The adversary has full access to the disk and can read and modify any data stored on it. When the TEE application calls \textsf{TEE.Unseal}, the adversary can freely choose which previously sealed data it will give to the TEE that will then decrypt and authenticate it. The adversary can also call all SCM operations through its interface. However, the adversary cannot directly manipulate the contents of the SCM beyond what the interface allows.

\subsection{Rollback Protection Algorithms}

We define two simple data structures. The first is a hash chain called $\textsf{history}_n = \textsf{hash}(\textsf{history}_{n-1} \parallel \textsf{data}_n)$ that is extended by concatenating the new data element to the previous chain value and then hashing the result. The second data structure is a tuple \textsf{state = \{model, budget, seed, pwd, history\}} that encodes the TEE implementation's persistent state, which includes the trained ML model \textsf{model}, the privacy budget \textsf{budget}, a randomness seed \textsf{seed}, an authorization password \textsf{pwd}, and the hash chain \textsf{history} that encodes the history of all execution inputs. These data structures are used by the following three algorithms.

\begin{algorithm}[t]
    \small
    \caption{\textsc{TEE-Init}. Initializes the rollback protection system.\label{fig:TEE-Init}}
    \SetKwFunction{SCMAuth}{SCM.Authorize}
    \SetKwFunction{SCMWrite}{SCM.Write}
    \SetKwFunction{TEESeal}{TEE.Seal}
    receive rand from data owners\;
    randomness $\gets$ XOR each rand\;
    derive seed and pwd from randomness\;
    \SCMAuth{pwd}\;
    \BlankLine
    history $\gets$ 0\;
    state $\gets$ $\{$empty model, full budget, seed, pwd, history$\}$\;
    \TEESeal{STATE, state}\;
    \SCMWrite{history, pwd}\;
\end{algorithm}

\subsubsection*{TEE-Init} Algorithm~\ref{fig:TEE-Init}, called \textsf{TEE-Init}, is used to initialize the rollback protection system. First, the TEE receives randomness shares from all data owners. To protect their data, honest data owners will send such randomness only once and refuse to do it again after they have sent training data to the TEE. The TEE algorithm combines the received randomness shares and derives two values, \textsf{seed} and \textsf{pwd}, using the combined randomness. Using the derived password, the TEE authorizes the use of the SCM. The algorithm also sets the current history to value zero and initializes the TEE's persistent state with an empty ML model, full privacy budget, the derived randomness seed and password, and an empty history as freshness tag (hash chain). Finally, the TEE seals its persistent state on disk and writes the current history value as the initial freshness tag to the SCM.

\begin{algorithm}[t]
    \small
    \caption{\textsc{TEE-Main}. Performs ML model training.\label{fig:TEE-Main}}
    \SetKwFunction{SCMAuth}{SCM.Authorize}
    \SetKwFunction{SCMWrite}{SCM.Write}
    \SetKwFunction{SCMRead}{SCM.Read}
    \SetKwFunction{TEESeal}{TEE.Seal}
    \SetKwFunction{TEEUnSeal}{TEE.Unseal}
    \SetKwFunction{hash}{hash}
    \SetKwFunction{train}{train}
    receive data from data owners\;
    state $\gets$ \TEEUnSeal{STATE} \KwSty{or} abort\;
    history $\gets$ \SCMRead{state.pwd}\;
    check (history = state.history) \KwSty{or} abort\;
    \BlankLine
    \TEESeal{DATA, data}\;
    history $\gets$ \hash{history $\parallel$ data}\;
    \SCMWrite{history, state.pwd}\;
    \BlankLine
    model $\gets$ \train{state.model, data, state.seed}\;
    update privacy budget\;
    state $\gets$ $\{$model, budget, state.seed, state.pwd, history$\}$\;
    \TEESeal{STATE, state}\;
\end{algorithm}

\subsubsection*{TEE-Main} Algorithm~\ref{fig:TEE-Main}, \textsf{TEE-Main}, is used for receiving new training data and re-training the model. A similar algorithm could be used to answer model queries from analysts. The TEE first receives training data from data owners and unseals its persistent state from disk. The execution can only proceed if the history (hash chain) value included in the unsealed persistent state matches the one in the SCM. If this is the case, the operation saves the input data for possible later use by sealing it and extends the history hash chain with the input data. The extended history chain value is written to the SCM. After that, the model can be re-trained, and the privacy budget updated. Finally, the TEE seals its current persistent state to disk.

\begin{algorithm}[t]
    \small
    \caption{\textsc{TEE-Recover}. Recovers from crashes.\label{fig:TEE-Recover}}
    \SetKwFunction{SCMAuth}{SCM.Authorize}
    \SetKwFunction{SCMWrite}{SCM.Write}
    \SetKwFunction{SCMRead}{SCM.Read}
    \SetKwFunction{TEESeal}{TEE.Seal}
    \SetKwFunction{TEEUnSeal}{TEE.Unseal}
    \SetKwFunction{hash}{hash}
    \SetKwFunction{train}{train}
    state $\gets$ \TEEUnSeal{STATE}\;
    data $\gets$ \TEEUnSeal{DATA}\;
    \BlankLine
    history $\gets$ \SCMRead{state.pwd}\;
    check (history = \hash{state.history $\parallel$ data}) \KwSty{or} abort\;
    \BlankLine
    model $\gets$ \train{state.model, data, state.seed}\;
    update privacy budget\;
    state $\gets$ $\{$model, budget, state.seed, state.pwd, history$\}$\;
    \TEESeal{STATE, state}\;
\end{algorithm}

\subsubsection*{TEE-Recover} Algorithm~\ref{fig:TEE-Recover}, \textsf{TEE-Recover}, is used for safe recovery, in case the TEE crashes during model training in \textsf{TEE-Main}. This operation unseals the previous state and input data, extends the history, and checks that the extended history matches the tag value in NVRAM. If this is the case, the model training and state sealing can proceed as before.

\subsubsection{Implementation}
\label{sec:rollback-implementation-details}
We implement our rollback protection using TPM 2.0 as an SCM. Messages from the TPM are sent in an audited session where the digest of the session is then signed, and a nonce is added to ensure authenticity and freshness. For sealing of data, we utilize 256-bit AES-GCM provided by the libsodium \cite{libsodium} C library. For hashing, we utilize SHA256.

\subsection{Security Analysis}

We provide an informal reasoning why the main security guarantees, privacy budget integrity, and timing side-channel minimization, hold for our solution.

\subsubsection*{Budget integrity} Since TEEs provide execution integrity, the adversary cannot manipulate the TEE's execution directly and, for example, remove the privacy budget enforcement check. However, the adversary can freely choose which version of the sealed state it provides to the TEE upon unsealing. In addition, the adversary can attempt to manipulate the freshness tag that is stored in the SCM. To cover both cases, we split our reasoning into two parts.

\emph{Part 1:} With respect to the sealed data, the adversary has five options. First, the adversary can provide the latest version of the sealed state (we call it version $n$) to the TEE. In this case, the TEE will accept the seal and enforce the budget check based on the correct value. Second, the adversary can provide the previous sealed state (version $n-1$). In this case, the \textsf{TEE-Main} routine will fail, but the \textsf{TEE-Recover} routine will accept the previous seal, assuming that the also sealed input data is the latest. The same DP computation will be executed again, starting from the previous budget value and resulting in the correctly updated budget value. No privacy budget re-use is possible. Third, the adversary can provide an older sealed state (version $< n-1$). In such cases, both \textsf{TEE-Main} and \textsf{TEE-Recover} will abort. 
Fourth, the adversary can provide no sealed state at all in an attempt to trick the TEE into believing that the system has not yet been initialized. However, in this case \textsf{TEE-Init} will fail, as long as at least one of the data owners is honest and does not send a randomness share again to the TEE. Fifth, the adversary can provide a fake sealed data element, with freely chosen content, that was not previously produced by the TEE. Both \textsf{TEE-Main} and \textsf{TEE-Recover} will refuse to accept such data, because the \textsf{TEE.Unseal} operation will return \textsf{fail} and the TEE will detect the manipulation and abort its execution. We conclude that none of the five options enables privacy budget reuse for the adversary.

\emph{Part 2:} Regarding the SCM, the adversary has two alternatives. First, the adversary can directly call \textsf{SCM.Write}, but since the adversary does not know the correct authorization password, this call will fail. Second, the adversary can call \textsf{SCM.Authorize} with an adversary-chosen password followed by \textsf{SCM.Write} with the same password and adversary-chosen data. This strategy will allow the adversary to change the SCM content, but the next time the TEE will call \textsf{SCM.Read}, the operation will return \textsf{fail} (due to the changed password) and the TEE will detect manipulation and abort. Thus, neither option enables the reuse of the privacy budget. 

Given \emph{Parts 1} and \emph{2}, we conclude that the adversary cannot reuse privacy budget, and thus budget integrity holds.

\subsubsection*{Timing Side-channel minimization} We say a rollback protection minimizes timing side-channels if the number of timing side-channel traces the adversary can collect is limited by the legitimate executions defined by the available privacy budget. The adversary can trigger DP computations using two TEE algorithms. First, the adversary can call \textsf{TEE-Main}, which checks the unsealed freshness tag against the SCM and writes the new freshness tag to the SCM \emph{before} the DP computation. This approach eliminates the side-channel oracle that is present in ElephantDP, for example, and ensures that every timing side-channel trace collected through \textsf{TEE-Main} will extend the history (freshness tag) and update the privacy budget that is available for the next DP computation. 

Second, the adversary can call \textsf{TEE-Recover}, which may reexecute the previous DP computation. Assuming that the hash function used in the hash chain is collision-resistant, the freshness tag check will only succeed if the previous state and the input data are the same as in the previous DP computation. Our solution ensures that the seed used to derive all pseudo-random values used in the DP computation is the same. Furthermore, our implementation hardening ensures that there is no leakage of sensitive data from arithmetic operations and similar instructions that are known to cause leakage (see Section~\ref{sec:hardening}). Together, these protections ensure that the only timing side-channel trace that the adversary can collect through \textsf{TEE-Recover} is the same trace that the adversary already has. We conclude that the adversary cannot collect more timing side-channel traces than what is permitted by the privacy budget.

\section{Timing Privacy}
\label{sec:timing_privacy}

We prove so-called Timing Privacy \cite{timingprivacy} in an idealized RAM model to validate our construction of \Cref{alg:mammothdptimingprivate}.

Timing Privacy (cf. \cref{def:timingprivacy}) is a framework for ensuring DP in the presence of timing side-channels. Intuitively, timing private mechanisms should not leak much more information about their input than what is already revealed by their output distribution. The formal definition is as follows.

\begin{definition}[Timing Privacy, special case of Definition 31 in \cite{timingprivacy}] \label{def:timingprivacy}
    A randomized mechanism $M\colon \mathcal{X} \times \mathcal{E} \mapsto \mathcal{R} \times \mathcal{E}$ satisfies $(1 \mapsto \rho(\alpha))$-Timing Privacy if, for any two neighboring datasets $X \sim X'$, any $\text{env}, \text{env}' \in \mathcal{E}$ and any $y \in \text{supp}(\text{out}(M(X, \text{env}))) \cap \text{supp}(\text{out}(M(X', \text{env}')))$:
    \begin{align}
    \nonumber \operatorname{D}_\alpha(&T_M(X, \text{env})|_{\text{out}(M(X, \text{env}))=y}\parallel \\
    &T_M(X', \text{env}')|_{\text{out}(M(X', \text{env}'))=y}) \leq \rho(\alpha)
    \end{align}
    where $T_M(X, \text{env})$ is the runtime of $M$ on inputs $X$ \& runtime environment $\text{env}$ and $\text{out}(M(X, \text{env}))$ is the output of $M$.
\end{definition}

For our proof, we assume that all hyperparameters are baked into \Cref{alg:mammothdptimingprivate} and are not part of the input. We note that the DP-approximated input length $\tilde{l}$ (cf. \Cref{alg:mammothdptimingprivate}) can be reused instead of being recomputed for the DP initial score (cf. \Cref{alg:dpinitscore}).
We use the $\omega$-bit Word RAM model's oracle \texttt{RAND} to sample uniform noise. We extend the model by oracles for selection between two values $u,v$ based on a boolean selector $b$ (cf. \Cref{alg:constanttimeselection}), logarithm, binomial coefficient, exponentiation, Bernoulli sampling, and Gaussian sampling. We assume all oracles to perform in a single time step, but note that this assumption is met in practice by making all oracles constant-time using the libftfp \cite{libfixedtimefixedpoint} and FACCT \cite{facct} libraries.
We also assume that the sequence of memory cells is large enough to store $\tilde{l}$ data points and that reading from uninitialized memory is safe, so we can read placeholder data points from uninitialized memory after the actual data points. We assume that one operation (e.g., calling an oracle or performing a multiplication) can be done per time step. Under these assumptions, we show the following result.

\begin{theorem}\label{thm:mammothtimingprivate}
    \Cref{alg:mammothdptimingprivate} satisfies $(1 \mapsto \rho_\text{input}(\alpha))$-Timing Privacy.
\end{theorem}

\begin{proof}
Let $X \sim X'$ be two neighboring datasets and $y \in \mathcal{R}$ an arbitrary output.
Let $T$ be the number of trees and $d$ the depth of trees.
Let $\tilde l \coloneqq |X| + \mathcal{N}(0, \sigma_\text{input}^2) + 2\sigma_\text{input}$ be the DP approximated input length of input $X$ (cf. \Cref{alg:mammothdptimingprivate}).

By direct analysis, counting one time step per oracle call and per arithmetic operation and treating expressions that depend only on hyperparameters as precomputed, we get the amount of operations for each component of \Cref{alg:ct-sbdt}:
the DP initial score in \Cref{alg:dpinitscore} ($3\tilde l + 12$),
and, in each round $t \in \{1, \dots, T\}$:
generating uniformly random splits for the random tree ($2^{d+1} - 1$),
updating and filtering individual privacy budgets at a fixed RDP order ($35\tilde l$),
Poisson subsampling in \Cref{alg:poissonsubsample} ($2\tilde l$),
initializing the leaf arrays ($2^d$),
ensemble prediction in \Cref{alg:predictensemble} for every data point ($\tilde l$, plus $\tilde l \cdot (6 \cdot 2^d - 2)$ for each of the $t-1$ previously trained trees, since the constant-time tree prediction in \Cref{alg:predicttree} visits every node),
computing gradients and Hessians in \Cref{alg:computegradientshessians} ($2\tilde l$),
building the leaf bitmasks in \Cref{alg:maketree} ($\tilde l \cdot (4 \cdot 2^d - 3) + 2^d$),
DP approximating the $2^d$ leaf values in \Cref{alg:dpleaf} ($2^d \cdot (10\tilde l + 7)$),
and writing the tree back to memory ($2^{d+1} - 1$).
Summing over all rounds and using $\sum_{t=1}^{T} (t-1) = (T^2 - T)/2$, we obtain a constant amount of operations independent of the input length
\begin{align}
    K = 12 + (13 \cdot 2^d - 2)\, T
\end{align}
and a constant amount of iterations for each data point in the input
\begin{align}
    k = 3 + (11 \cdot 2^d + 38)\, T + (3 \cdot 2^d - 1)\, T^2.
\end{align}
Let $M$ be the mechanism of \Cref{alg:sbdt_skeleton} on the padded input. The runtimes yield
\begin{align}
&\nonumber\phantom{~=} T_M(X, \text{env})|_{\text{out}(M(X,\text{env}))=y} \\
&\nonumber= \big(\mathcal{N}(|X|, \sigma_\text{input}^2) + 2\sigma_\text{input}\big) \cdot k + K \\
&= \mathcal{N}\big((|X| + 2\sigma_\text{input}) \cdot k + K,\; k^2 \sigma_\text{input}^2\big)
\end{align}
and similarly for $X'$. We compute the Timing Privacy guarantee of $M$:
\begin{align}
\nonumber D_\alpha\big(&T_M(X, \text{env})|_{\text{out}(M(X,\text{env}))=y} \parallel \\
&\nonumber T_M(X', \text{env}')|_{\text{out}(M(X',\text{env}'))=y}\big) \\
\nonumber= D_\alpha\big(&\mathcal{N}((|X| + 2\sigma_\text{input}) k + K, k^2 \sigma_\text{input}^2) \parallel \\
&\nonumber \mathcal{N}((|X'| + 2\sigma_\text{input}) k + K, k^2 \sigma_\text{input}^2)\big) \\
\nonumber= D_\alpha\big(&\mathcal{N}(0, k^2 \sigma_\text{input}^2) \,\|\, \mathcal{N}(k, k^2 \sigma_\text{input}^2)\big)
\intertext{(using \cite[Proposition 7]{DBLP:journals/corr/Mironov17})}
\le\phantom{D_\alpha\big(}& \frac{\alpha}{2\sigma_\text{input}^2} \eqqcolon \rho_\text{input}(\alpha)
\end{align}
\end{proof}

\newpage
\clearpage

\clearpage
\onecolumn
\section{Pseudocode for Constant-Time S-BDT}
\label{sec:ctsbdtpseudocode}

\begin{algorithm}[H]
    \small
    \newcommand{\single}[1]{\texttt{TrainSingleTree}}
    \caption{Constant-time \textsc{S-BDT}.\label{alg:ct-sbdt}}
    \SetKwFunction{RandomBinaryTree}{RandomBinaryTree}
    \SetKwFunction{PrivacyFilter}{PrivacyFilter}
    \SetKwFunction{PoissonSubsample}{PoissonSubsample}
    \SetKwFunction{NoisyAverageLabel}{NoisyAverageLabel}
    \SetKwFunction{NoisyAverageGradient}{NoisyAverageGradient}
    \SetKwFunction{ComputeGradientsHessians}{ComputeGradientsHessians}
    \SetKwFunction{NoisyAverageHessian}{NoisyAverageHessian}
    \SetKwFunction{NoisyClipGradSum}{NoisyClipGradSum}
    \SetKwFunction{NoisyClipHessSum}{NoisyClipHessSum}
    \SetKwFunction{UpdateGradient}{UpdateGradient}
    \SetKwFunction{UpdateHessian}{UpdateHessian}
    \SetKwFunction{DPInitScore}{DPInitScore}
    \SetKwFunction{PredictTree}{PredictTree}
    \SetKwFunction{DPLeaf}{DPLeaf}
    \SetKwFunction{Bitmask}{Bitmask}
    \SetKwFunction{MakeTree}{MakeTree}
    \KwIn{$T$: number of rounds, $D$: training data}
    $f^{(0)} = \DPInitScore{\dots}$\tcp*{cf. \Cref{alg:dpinitscore}}
    $f = (f^{(0)})$\tcp*{ensemble, initiated with init score}
    \For(\tcp*[f]{train tree $f^{(t)}$}){$t = 1$ \KwSty{to} $T$}{
      $f^{(t)} =$ \RandomBinaryTree{$\dots$}\;
      $\Bitmask_F =$ \PrivacyFilter{$D, \dots$}\; 
      $\Bitmask_S =$ \PoissonSubsample{$D, \dots$}\;
      $\forall j\colon I^{(\text{train)}}_{\text{Leaf}_j} = []$\tcp*{initialize leaf array}
      $\mathcal{P} = \text{\PredictEnsemble{$f,D$}}$\tcp*{cf. \Cref{alg:predictensemble}}
      $(g_i, h_i)_{i=1}^{|D|}$ = \ComputeGradientsHessians{$\mathcal{P}, \dots$}\;
      \MakeTree{\dots}\;
      \For(\tcp*[f]{cf. \Cref{eq:leaf_value}}){\KwSty{each} leaf $v_j$ \KwSty{in} $f^{(t)}$}{
        $f_j^{(t)}(\cdot) \coloneqq$ \DPLeaf{$\dots$}\;
      }
      Append tree $f^{(t)}$ to ensemble $f$\;
    }
    \Return $f$\;
\end{algorithm}

\begin{algorithm}[H]
    \small
    \caption{MakeTree\label{alg:maketree}}
    \SetKwFunction{Bitmask}{Bitmask}
    \SetKwFunction{PoissonSubsample}{PoissonSubsample}
    \KwIn{$D$: training data, $v$: current node, $\Bitmask_L$: leaf bitmask}
    \SetKwInOut{KwIn}{\color{white}\phantom{Input}}
    
    \If{$v$ is a leaf}{
        Store $\Bitmask_L$ in $v.\Bitmask_L$\;
        \Return\;
    } \Else{
              
        Copy $\Bitmask_L$ into $\Bitmask_L^{\text{left}}$\;
        Copy $\Bitmask_L$ into $\Bitmask_L^{\text{right}}$\;
        \For{\KwSty{each} sample $(x_i,y_i)$ \KwSty{in} $D$}{
            $\Bitmask_L^{\text{left}}[i] =  v_i.\text{split\_criterion}(x_i)$\;
            $\Bitmask_L^{\text{right}}[i] = 1 - v_i.\text{split\_criterion}(x_i)$\;
        }
        \MakeTree{$D, v.\text{child.left}, \Bitmask_L^{\text{left}}$}\;
        \MakeTree{$D, v.\text{child.right}, \Bitmask_L^{\text{right}}$}\;
    }
\end{algorithm}

\begin{algorithm}[H]
    \small
    \caption{DPInitScore: Compute a DP initial score \label{alg:dpinitscore}}
    \SetKwFunction{Gauss}{CTGauss}
    \SetKwFunction{Clamp}{CTClamp}
    \KwIn{$D$: training dataset, $m^*$: clipping bound on labels}
    \SetKwInOut{KwIn}{\color{white}\phantom{Input}}
    \KwIn{$\varepsilon_\text{init}$: DP privacy budget for initial score}
    \KwIn{$\eps_{\text{ds}} = 0.025$: DP privacy budget for dataset size}
    \BlankLine
    $|D|_{\text{priv}} = 0$\;
    $M = 0$\;
    \For{\KwSty{each} sample $(x_i,y_i)$ \KwSty{in} $D$}{
        $|D|_{\text{priv}} = |D|_{\text{priv}} + 1$ \label{dpreleaseds} \;
    }

    $|D|_{\text{priv}} = |D|_{\text{priv}} + \text{\Gauss{$0, 2\cdot\ln{}(1.25/\delta)/\eps_{\text{ds}}^2$}}$\tcp*{constant-time Gauss}
    
    \If{regression}{
        \For{\KwSty{each} sample $(x_i,y_i)$ \KwSty{in} $D$}{
            $M = M +$ \Clamp{$y_i, -m^*, m^*$} \;
        }
        $M_{\text{priv}} = \frac{1}{|D|_{\text{priv}}} \cdot M + \text{\Gauss{$0, 2\ln{}(1.25/\delta) (m^*/(|D|_{\text{priv}} \cdot \varepsilon_\text{init}))^2$}}$ \label{algdpinitscorelaplace}\;
        \Return $M_{\text{priv}}$
     }\ElseIf{classification}{
         \For{\KwSty{each} sample $(x_i,y_i)$ \KwSty{in} $D$}{
            $M = M +$ \Clamp{$y_i, 0, m^*$} \;
        }
        $M_{\text{priv}} = \frac{1}{|D|_{\text{priv}}} \cdot M + \text{\Gauss{$0, 2\ln{}(1.25/\delta) (m^*/(|D|_{\text{priv}} \cdot \varepsilon_\text{init}))^2$}}$\tcp*{constant-time Gauss}
         \Return $\ln(\sfrac{M_{\text{priv}}}{1 - M_{\text{priv}}})$
     }
\end{algorithm}

\begin{algorithm}[H]
    \small
    \caption{DPLeaf: Compute a DP leaf node \label{alg:dpleaf}}
    \SetKwFunction{Gauss}{CTGauss}
    \SetKwFunction{Clamp}{CTClamp}
    \SetKwFunction{max}{max}
    \KwIn{$D$: training data, $v$: leaf node}
    \SetKwInOut{KwIn}{\color{white}\phantom{Input}}
    \KwIn{$(g \in D, h \in D)$: gradient/Hessian of data points in $D$} 
    \KwIn{$(g^*, h^*)$: clipping bound on gradients and Hessians}
    \KwIn{$(r_{1}, r_{2})$: noise weights for leaf value}
    \KwIn{$\lambda, \beta$: regularization parameters}
    \KwIn{$\sigma_\text{leaf}^2$: unweighted variance of Gaussian for leaves}
    \KwIn{$\Bitmask_F$: privacy filter bitmask}
    \KwIn{$\Bitmask_S$: subsample bitmask}
    \For{\KwSty{each} sample $(x_i,y_i)$ \KwSty{in} $D$}{
        $w = w \, +$ \Clamp{$h_i, 0.0, h^*$} $\cdot \, v.\Bitmask_L[i] \cdot \Bitmask_F[i] \cdot \Bitmask_S[i]$\;
        $u = u \, +$ \Clamp{$g_i, -g^*, g^*$} $\cdot \, v.\Bitmask_L[i] \cdot \Bitmask_F[i] \cdot \Bitmask_S[i]$\;
    }
    $\tilde{w} = \text{$\lambda + w + \text{\Gauss{$0.0, (h^{*})^{2}\sigma_\text{leaf}^2 / (2 \cdot r_1)$}}$}$ \label{algdpleafnoisedsupport}\tcp*{constant-time Gauss}
    $\tilde{u} = u + \text{\Gauss{$0.0, (g^{*})^{2} \sigma_\text{leaf}^2/ (2 \cdot r_2)$}}$ \label{algdpleafnoisedsum}\tcp*{constant-time Gauss}
    \Return \Clamp{$v = \tilde{u} / \tilde{w}, -\beta, \beta$} \label{algdpleafvalue}
\end{algorithm}

\begin{algorithm}[H]
    \small
    \caption{CTClamp: Constant-time clamp}
    \SetKwFunction{Gauss}{Gauss}
    \SetKwFunction{Clamp}{Clamp}
    \SetKwFunction{max}{max}
    \KwIn{$v$: value to clamp, $l$: lower bound, $r$: upper bound}
    $v' = l \cdot (v < l) + v \cdot (v \geq l)$\;
    $v' = l \cdot (v' > r) + v' \cdot (v' \leq r)$\;
    \Return v'\;
\end{algorithm}

\begin{algorithm}[H]
    \small
    \caption{PredictEnsemble \label{alg:predictensemble}}
    \SetKwFunction{Length}{Length}
    \SetKwFunction{PoissonSubsample}{PoissonSubsample}
    \SetKwFunction{NoisyAverageLabel}{NoisyAverageLabel}
    \SetKwFunction{NoisyAverageGradient}{NoisyAverageGradient}
    \SetKwFunction{NoisyAverageHessian}{NoisyAverageHessian}
    \SetKwFunction{UpdateGradient}{UpdateGradient}
    \SetKwFunction{UpdateHessian}{UpdateHessian}
    \SetKwFunction{DPInitScore}{DPInitScore}
    \SetKwFunction{LeafIndexOf}{LeafIndexOf}
    \KwIn{$x$: data point, $f$: S-BDT ensemble, $\eta$: learning rate}
        $\hat{y} = f^{(0)}$\;
        \For{$t = 1$ \KwSty{to} $\Length{f}-1$}{
            $v = f^{(t)}.\text{root}$\;
            $\hat{y} = \hat{y} + \eta \cdot \text{\PredictTree{$v, x$}}$\;
        }
    \Return $\hat{y}$\;
\end{algorithm}

\begin{algorithm}[H]
    \small
    \caption{PredictTree: Compute prediction of tree \label{alg:predicttree}}
    \SetKwFunction{Gauss}{Gauss}
    \SetKwFunction{Clamp}{Clamp}
    \SetKwFunction{PredictTree}{PredictTree}
    \SetKwFunction{max}{max}
    \KwIn{$v$: node of tree, $x$: data point}

    \If{$v$ is leaf}{
        \Return prediction of $v$ \\
    }
    $\text{l} = \text{\PredictTree}(v.\text{child}.\text{left}, x)$ \\
    $\text{r} = \text{\PredictTree}(v.\text{child}.\text{right}, x)$ \\

    \Return $v.\text{split\_criterion}(x) \cdot l +  (1-v.\text{split\_criterion}(x)) \cdot r$
\end{algorithm}

\begin{algorithm}[H]
    \small
    \caption{ComputeGradientsHessians: Compute gradients and Hessians \label{alg:computegradientshessians}}
    \SetKwFunction{Gauss}{Gauss}
    \SetKwFunction{Clamp}{Clamp}
    \SetKwFunction{max}{max}
    \KwIn{$\mathcal{P}$: ensemble prediction on $D$, $D$: data}
    \If{regression}{
        \For {\KwSty{each} sample $(x_i, y_i)$ \KwSty{in} $D$}{
            $g_i = \mathcal{P}[i] - y_i$ \\
            $h_i = 1$ \\
        }
     }\ElseIf{classification}{
        \For {\KwSty{each} sample $(x_i, y_i)$ \KwSty{in} $D$}{
            $g_i =  1 / (1 + \exp{}(-\mathcal{P}[i])) - y_i$ \\
            $h_i = (y_i + g_i) \cdot (1 - y_i - g_i)$ \\
        }
     }
     \Return $(g_i, h_i)_{i=1}^{|D|}$
\end{algorithm}

\begin{algorithm}[H]
    \small
    \caption{PoissonSubsample: Generate subsample bitmask\label{alg:poissonsubsample}}
    \SetKwFunction{Gauss}{Gauss}
    \SetKwFunction{Clamp}{Clamp}
    \SetKwFunction{Bitmask}{Bitmask}
    \SetKwFunction{BernoulliBit}{CTBernoulliBit}
    \SetKwFunction{max}{max}
    \KwIn{$D$: data, $\gamma$: subsampling ratio}
    $\Bitmask_S = [0, ..., 0]$\;
    \For {\KwSty{each} sample $(x_i, y_i)$ \KwSty{in} $D$}{
        $\Bitmask_S[i] = \text{\BernoulliBit{}}$\tcp*{constant-time Bernoulli}
    }
    \Return $\Bitmask_S$ \;
\end{algorithm}

\end{document}